\documentclass[10pt,fleqn,final,5p,times,twocolumn,sort&compress,lefttitle]{elsarticle}

\usepackage{lineno,hyperref}
\usepackage{amsmath}
\usepackage{color}
\modulolinenumbers[5]
\usepackage{fancyhdr}

\usepackage[labelfont=bf,labelsep=period]{caption}

\journal{Journal of Nuclear Materials}

\begin{document}

\begin{frontmatter}

\title{\LARGE Characterizing accelerated precipitation in proton irradiated steel}

\author[UoB,UoW]{\large Mark Laver \corref{corauthor}}
\cortext[corauthor]{Corresponding author.}

\author[UoM]{Brian J.\ Connolly}

\author[UoBb]{Christopher Cooper}

\author[PSI]{Joachim Kohlbrecher}

\author[PSI]{Stavros Samothrakitis}

\author[RR]{Keith Wilford}

\address[UoB]{School of Physics and Astronomy, University of Birmingham, Birmingham B15 2TT, United Kingdom}
\address[UoW]{Department of Physics, University of Warwick, Coventry CV4 7AL, United Kingdom}
\address[UoM]{Department of Materials, University of Manchester, Oxford Road, Manchester M13 9PL, United Kingdom}
\address[UoBb]{School of Metallurgy and Materials, University of Birmingham, Birmingham B15 2TT, United Kingdom}
\address[PSI]{Paul Scherrer Institute, 5232 Villigen PSI, Switzerland}
\address[RR]{Rolls-Royce, PO Box 2000, Raynesway, Derby DE21 7XX, United Kingdom \vspace{-0.8cm}}

\begin{abstract}
Ion irradiation provides a promising substitute to neutron tests for investigating the effects of radiation on materials for fission and fusion reactor plants.
Here we show proton irradiation can quantitatively reproduce precipitation that leads to embrittlement in reactor pressure vessel steels, at dose rates 10$^4$ times greater than experienced in fission reactor operation.
Small-angle neutron scattering (SANS) is used to characterize precipitate size distributions in copper-containing steels irradiated to average doses of $\simeq 7$\,mdpa with 5\,MeV protons.
Comparing our results with the literature on reactor pressure vessel steels containing $\geq 1$\,at.\% nickel,
we find a power-law scaling of dose with exponent 0.25--0.30 accounts for the effects of dose rate on precipitate volume fraction
over 6 orders of magnitude in dose rate. In conjunction with dose rate, carbon is identified as performing a leading role in determining precipitate sizes,
adding to the known effects of nickel, manganese and irradiation temperature.
We discuss the composition of precipitates inferred from SANS, taking previous atom probe tomography studies into consideration.
\end{abstract}

\begin{keyword}
Precipitation kinetics \sep Irradiation-hardening rate \sep Ion irradiation \sep Nanostructure
\sep Small-angle scattering
\end{keyword}

\end{frontmatter}

\section{Introduction}
\label{sec:introduction}
The irradiation of materials with high energy particles spawns a profusion of microscopic effects.
These include formation of dislocation loops, vacancy-solute clusters, voids and bubbles;
nucleation or dissolution of precipitates and secondary phases;
acceleration of corrosion; segregation of solute atoms to or from surfaces, grain boundaries, dislocations or voids~\cite{Okamoto79, WasBook}.
These effects may be radiation {\em enhanced}, for example the acceleration of precipitation due to enhanced diffusion~\cite{WasBook, Odette83}
or radiation {\em induced}, for example precipitation in under-saturated solid solutions~\cite{Meslin13,Belkacemi18}.
The relative significance of each effect depends sensitively on irradiation conditions (temperature, corrosive environment,
incident particle, its energy, flux and fluence) as well as on material (composition, microstructure).
Indeed the sensitivity to these variables presents a hurdle in designing experiments to stay ahead of arising challenges in radiation materials science.

A case in point is the neutron embrittlement of reactor pressure vessel (RPV) steels~\cite{VanDuysen17, Williams19, Odette19, Hyde15}
where the realized importance of neutron flux effects has led to regulatory models being constructed from surveillance data only~\cite{Williams19}
and to restrictions on the permissible lead factors of surveillance specimens in regulatory codes~\cite{ASTM16}.
This does not imply that accelerated irradiation studies are wanting, rather the opposite: they are
needed to explore and understand new phenomena identified by surveillance programmes in operational plants
and to develop new materials for next generation fission and future fusion reactors.
Here we use small-angle neutron scattering (SANS) to characterize the nanostructure of proton-irradiated steel specimens
and show that 5\,MeV proton irradiations with lead factors of $10^4$
can emulate fast neutron radiation damage in RPV steels.

Accelerated irradiation experiments and experiments with surrogate irradiating particles are most valuable when
they quantitatively reproduce the prominent microscopic effects.
To compare radiation damage caused by different particles, it is useful to quantify radiation dose in terms of displacements per atom (dpa).
For example, the typical $E > 1$\,MeV neutron flux on the inner side of an operating pressurized water reactor RPV is $\approx 6 \times 10^{14}$\,n\,m$^{-2}$\,s$^{-1}$~\cite{VanDuysen17}
which, using the dpa cross-section $\sigma_{\mathrm{dpa}} = 1.5 \times 10^{-25}$\,m$^2$~\cite{Mathon97},
corresponds to a dose rate of $9 \times 10^{-11}$\,dpa\,s$^{-1}$ or $\simeq 3$\,mdpa/yr,
and after 35 service years the fluence is $\approx 7 \times 10^{23}$\,n\,m$^{-2}$ or 0.1\,dpa.
MeV neutrons incident on steel produce large displacement cascades with $\approx 100$\,keV of energy transferred to lattice atoms.
Interstitial clustering within these cascades is significant and drives nucleation of dislocation loops~\cite{Gan01}.
Protons or heavier ions also produce cascades and so are appropriate as surrogates for neutrons.
MeV protons produce smaller cascades than MeV neutrons, with a few hundred eV of energy transferred on average by the Coulomb interaction.
Despite differences in energy transfer and cascade size, the net microscopic damage i.e.\ interstitial clustering within cascades and
dislocation loop densities and sizes, is found to be {\em roughly} equivalent
since the differences are counterbalanced by more numerous proton cascades and by
greater damage efficiency due to reduced recombination in smaller proton cascades~\cite{WasBook,Gan01,Was15}.
A shift in irradiation temperature may further improve agreement between neutron and surrogate ion irradiations by compensating for dose rate effects~\cite{Mansur78,Okamoto79,WasBook,Was15}.
The optimal shift depends on the reference temperature and the specific microscopic effect to be reproduced. For example, a recent dual Fe$^{2+}$ and He$^{2+}$
irradiation study of T91 ferritic-martensitic steel finds a $+60$ to $70^\circ$C shift accurately
reproduces dislocation and cavity microstructures of fast neutron irradiations at reference temperatures 370$^\circ$C--420$^\circ$C, but the ion irradiations
are unable to reproduce grain boundary segregation or Ni/Si-rich cluster formation effects~\cite{Taller19}.
Further examples of the use of protons and  ions as surrogates for neutrons are reviewed in Ref.~\cite{Was15}.
Studies on austenitic stainless steels published since this review
provide further demonstrations that neutron radiation-induced segregation at light-water reactor operating temperatures $\simeq 300^\circ$C,
leading to both hardening and susceptibility to irradiation-assisted stress corrosion cracking,
can be emulated by proton irradiations with a $+60^\circ$C temperature shift~\cite{Stephenson15,Song18,Jin19}.

\begin{figure}
\begin{center}
\includegraphics[width=6.4cm]{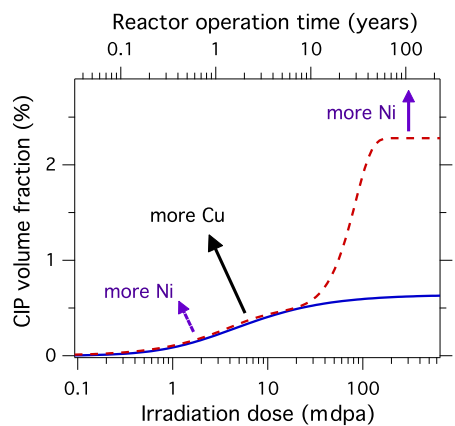}
\end{center}
\caption{\label{fig:schematic} Schematic illustrating growth of copper-initiated precipitates (CIPs) under irradiation.
Approximate axis scales are indicated for reactor pressure vessel (RPV) steels containing high ($\simeq 0.3$\,at.\%) Cu and high ($\gtrsim 1$\,at.\%) Ni, such as the steel studied in this work.
Top axis indicates typical service years for the inner side of a RPV.
Arrows show leading effects of Cu and Ni content.
Solid line depicts the empirical curve typically used in regulatory models (see e.g.\ \cite{Williams19} or \cite{Debarberis05}).
Dashed line depicts double-Avrami model curves where one component embodies ``late-blooming'' at high doses~\cite{Odette19}.}
\end{figure}
In this work we consider low alloy ferritic/bainitic steels that have particular application in light-water RPVs.
The effects of radiation on these materials are comprehensively covered in several recent reviews~\cite{VanDuysen17, Williams19, Odette19, Hyde15}.
Here we focus our treatment on copper-containing steels in which the formation of copper-initiated precipitates (CIPs) is the dominant microscopic effect that
causes embrittlement and determines reactor service lifetimes.
With increasing service life i.e.\ neutron fluence, the volume fraction of CIPs increases as they form and grow, then starts to plateau
as matrix Cu is depleted (Fig.~\ref{fig:schematic}).
Embrittlement, as measured by a positive shift in ductile-to-brittle transition temperature or an increase in yield stress,
correlates well with the square root of precipitate volume fraction~\cite{Williams19, Fisher85, Wagner16, Dohi10, Fujii13, Wells14, Almirall19}.
This relationship follows the ideas of Russell-Brown precipitate hardening~\cite{Russell72,Williams01},
though more refined hardening models may better describe experimental data~\cite{Wagner16}.
Density functional theory and Monte Carlo calculations show that the diffusivity of copper in bcc iron is dominated by vacancies, which drag copper atoms to point defect
sinks~\cite{Vincent05, Barashev06, Messina14},
so the first step in modelling radiation-accelerated precipitation kinetics is to account for
excess vacancies induced by radiation and the resulting radiation-enhanced diffusion of Cu~\cite{Sizmann78,Odette83,Fisher85,Odette05}.
Interstitial clusters and prismatic dislocation loops generated in displacement cascades also interact strongly with Cu and other solute atoms in RPV steels,
namely P, Si, Mn and Ni, as recently overviewed in Ref.~\cite{Castin20}.
These solutes are expected to enrich at defect sinks during irradiation at RPV operating temperatures $\simeq 300^\circ$C,
with the diffusion of Cu, Ni and Si led by the vacancy drag mechanism and for P and Mn by the dumbbell mechanism~\cite{Messina20}.
It is well established in experiments that Ni, Mn and Si~\cite{Miller87,Miller92,Monzen04,Isheim06,Hyde11,Styman15,Styman18} enrich at the CIP--matrix interfaces.
As the Ni, Mn and Si contents are usually higher than the impurity Cu content in RPV steels
and the kinetics of their radiation-induced segregation at CIPs are slower than initial CIP growth due to Cu,
these solutes accumulate heterogeneously at CIPs long after the matrix is depleted of Cu.
This `after-effect' increase of precipitate volume fraction gives rise to concomitant hardening at high fluences, dubbed ``late-blooming''~\cite{Odette19,Wells14,Mamivand19,Almirall20}
(cf.\ Fig.~\ref{fig:schematic}).
The driving mechanism behind the clustering of Ni, Mn and Si in the absence of Cu is a question of current debate (see e.g. \cite{Castin20}).
For copper-containing steels, like those studied here with 0.3\,wt.\% Cu, the situation at modest doses is clear:
the Cu content lies above the low solubility limit (0.0003\,wt.\%~\cite{Salje77} to 0.05\,wt.\%~\cite{VanDuysen17},
from extrapolation of high temperature data to 300$^\circ$C)
so we expect Cu-led precipitation from the supersaturated solid solution in equilibrium conditions, although at 300$^\circ$C the kinetics are slow.
Indeed CIPs with Ni, Mn and Si-enriched interfaces are observed in long-time thermal ageing experiments of RPV steels
at 405$^\circ$C with 0.3\,at.\% Cu after 16 months~\cite{Zelenty16}, at 330$^\circ$C with 0.4\,at.\% Cu after 11 years~\cite{Styman15,Styman18}
and at 345$^\circ$C with 0.1\,at.\% Cu after 28 years~\cite{Lindgren18}.
In view of their palpable thermodynamic stability at 300$^\circ$C
we use the term ``copper-initiated {\em precipitates}'' (CIPs) throughout to describe the solute clusters that develop in Cu-containing RPV steels.

Nickel has a synergistic effect with copper on the CIP hardening contribution, increasing both the height of the initial plateau
associated with Cu depletion from the matrix~\cite{Hawthorne82,Eason13,Solt93}
and also the hardening and CIP volume fraction in the limit of very high fluence~\cite{Almirall20} (Fig.~\ref{fig:schematic}).
This limit, as evoked by cluster dynamics modelling~\cite{Mamivand19}, lies at $\approx 1$\,dpa.
It is reached in a recent accelerated neutron and ion irradiation study that finds the volume fraction scales as $(2 \, \mathrm{Ni} + \mathrm{Cu})$
in RPV steels containing modest Mn or Si~\cite{Almirall20}.
Other recent studies on RPV steels also use surrogate particles,
namely keV or MeV protons~\cite{Was05,Shi19}, or MeV Fe ions~\cite{Murakami16,Roder18,Kedharnath19,Almirall20}.
For keV protons or MeV Fe ions, the damage depth is typically a few micrometres or less.
To characterize the radiation damage in this thin layer, studies draw upon contemporary nano-techniques such as
atom probe tomography~\cite{Almirall20} and nanoindentation~\cite{Shi19,Murakami16,Roder18,Kedharnath19,Liu14}.
An attribute of these techniques is that the minute volume sampled may not be representative of the bulk,
with significant variation arising from mesoscale heterogeneity~\cite{Murakami16}.
Here we use a conventional bulk technique --- small-angle neutron scattering (SANS) --- to characterize CIPs
in copper-containing steel specimens irradiated by 5\,MeV protons at 300$^\circ$C and at 400$^\circ$C.
SANS measures volume fractions and size distributions of CIPs averaged over macroscopic specimen volumes.
Comparing quantitatively our proton irradiations against neutron results in the literature,
including irradiations at neutron fluxes similar to RPV operational conditions~\cite{Bergner08,Hyde01,Eason07},
we find that 300$^\circ$C proton irradiation with a lead factor of $10^4$ is able to reproduce CIP formation and emulate in-service neutron irradiation of RPVs.

\section{Material and methods}
\subsection{Material}
The chemical composition of the split-melt bainitic model steel used in this study is shown in Table~\ref{tab:composition}.
It was supplied by Rolls-Royce as part of an effort to investigate high strength, low alloy steels~\cite{Almirall19, Jenkins20}. 
The as-received material is identical to the high Cu model steel used in the thermal ageing and atom
probe tomography study of Zelenty {\it et al.}~\cite{Zelenty16}, where details of the forging and processing can be found.
In brief, after forging and rolling to 25\,mm thickness,
the material was austenized at 920$^\circ$C for 1 hour followed by an air cool, then tempered at 600$^\circ$C for 5 hours followed by an air cool.
The yield strength of the unirradiated material is $\simeq 700$\,MPa.
Slab-shaped specimens of thickness $\simeq 0.2$\,mm were cut from the as-received material for proton irradiation and SANS experiments.

\begin{table*}
\caption{\label{tab:composition} Analysis of steel composition.}
\begin{center}
\begin{tabular}{rccccccccc}
\hline
      &  C   &  Si &   P   &  Cr &  Mn  &  Ni  &  Cu  &  Mo  & Fe \\ \hline
wt.\% & 0.21 & 0.2 & 0.008 & 0.1 & 1.48 & 3.47 & 0.31 & 0.52 & bal. \\
at.\% & 0.97 & 0.4 & 0.014 & 0.1 & 1.50 & 3.28 & 0.27 & 0.30 & bal. \\ \hline
\end{tabular}
\end{center}
\end{table*}

\begin{table}
\caption{\label{tab:irradiation} Summary of proton irradiation conditions. Dose and dose rate are averaged over the damaged volume shown in Fig.~\ref{fig:SRIM}.}
\begin{center}
\begin{tabular}{cccc}
\hline
$T_\mathrm{irr}$ & $E$ & Dose & Dose rate \\
($^\circ$C) & (MeV) & ($10^{-3}$\,dpa) & ($10^{-7}$\,dpa\,s$^{-1}$) \\ \hline
300 & 5 & 7.2 & 8.8 \\
400 & 5 & 7.2 & 7.4 \\ \hline
\end{tabular}
\end{center}
\end{table}

\subsection{Proton irradiation}
\label{sec:protonirr}
\begin{figure}
\begin{center}
\includegraphics[width=8.8cm]{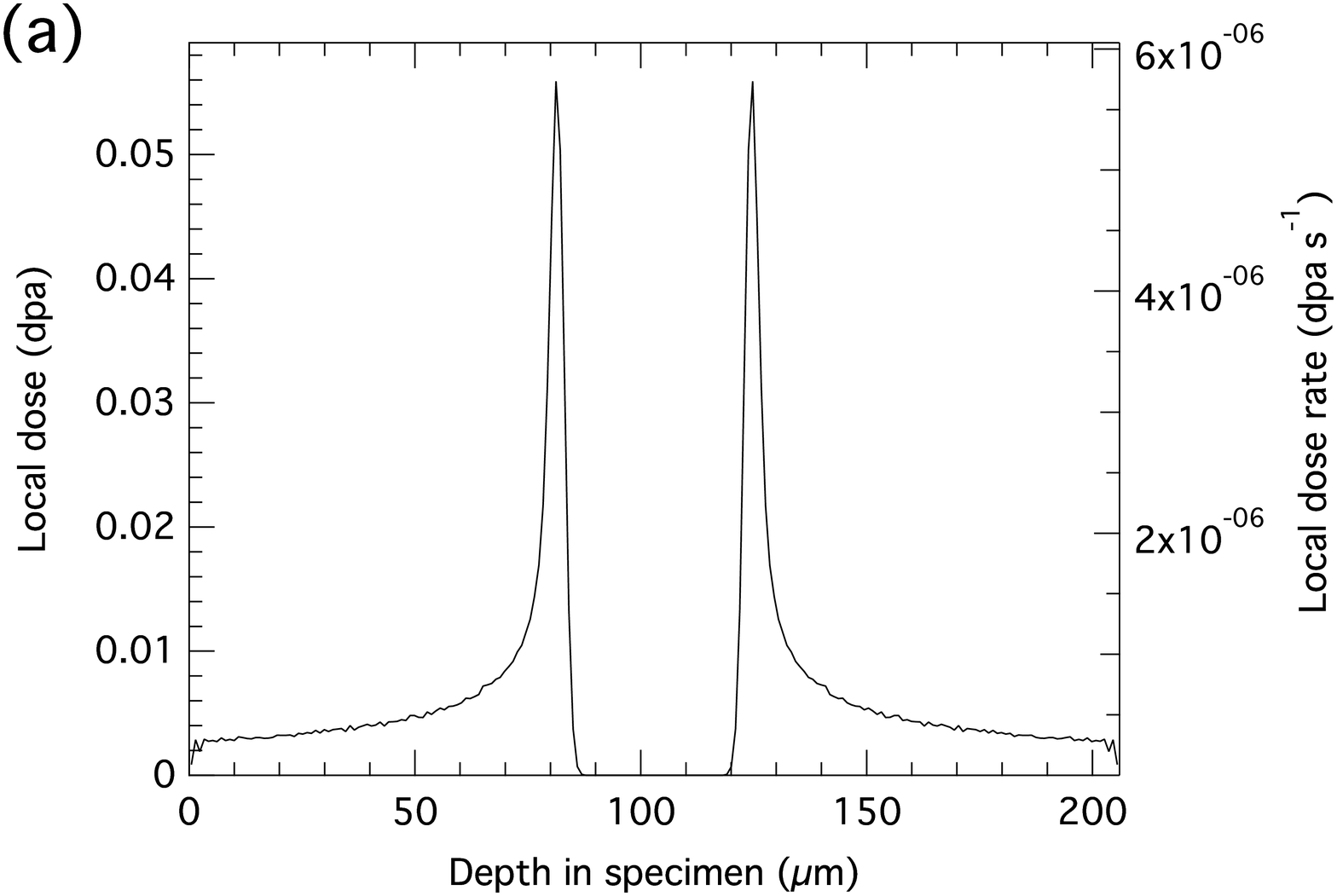}
\includegraphics[width=8.0cm]{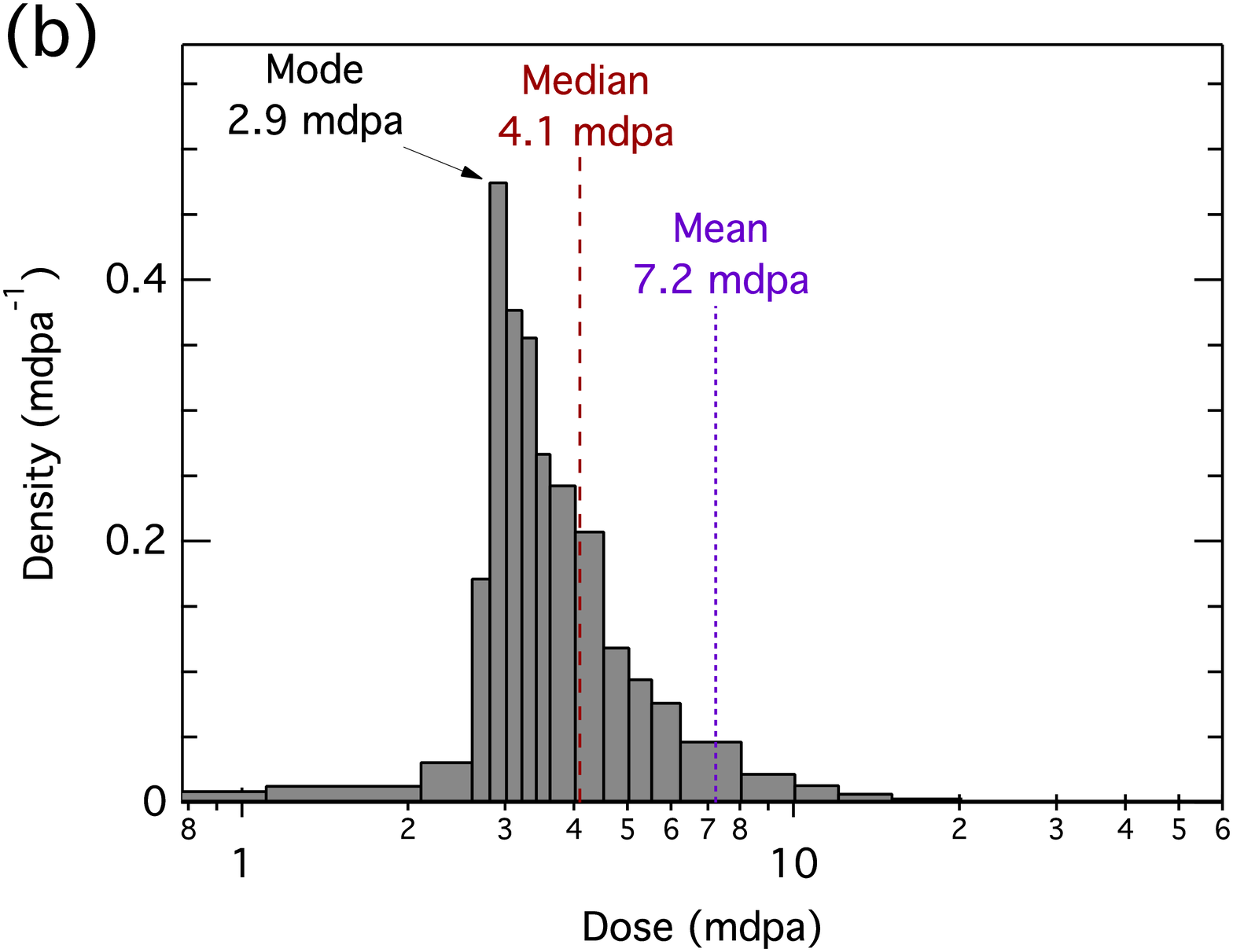}
\end{center}
\caption{\label{fig:SRIM} Proton damage in our model steels. (a) Damage profile through the thickness of our slab-shaped specimen irradiated by 5\,MeV protons at 400$^\circ$C, calculated using the SRIM Monte Carlo code.
Specimens were irradiated twice, once from each side, giving the two Bragg peaks shown.
Average dose is 6.1\,mdpa over the full thickness, or 7.2\,mdpa excluding the undamaged central region.
(b) Density histogram showing dose distribution in our specimens calculated from (a), excluding undamaged central region. Modal, median and mean doses are indicated.
Our specimen irradiated by 5\,MeV protons at 300$^\circ$C has a similar dose profile but a slightly higher dose rate (see text Sec.~\ref{sec:protonirr} and Table~\ref{tab:irradiation}).}
\end{figure}

The irradiation conditions are summarized in Table~\ref{tab:irradiation}.
Proton irradiations were carried out at the MC40 cyclotron facility at the University of Birmingham.
To control the irradiation temperature, specimens were mounted on a heater block in vacuum.
The first specimen was irradiated at 300$^\circ$C and the second at 400$^\circ$C, with the uncertainty in temperature estimated to be $\pm 5^\circ$C,
measured by a thermocouple in contact with each specimen.
Each specimen was irradiated twice, once from each side, with 5\,MeV protons
rastered over a 1.08\,cm$^2$ aperture placed close to the sample position upstream.
Each irradiation took $\approx 2\tfrac12$\,hours, with $2.5 \times 10^{17}$ protons deposited per specimen side.
Dose calculations were carried out using the `quick' Kinchin-Pease option in the SRIM Monte Carlo code with a threshold displacement energy of 40\,eV,
following the recommendations of ASTM E521 / Ref.~\cite{Stoller13}: the resulting damage profile is illustrated in Fig.~\ref{fig:SRIM}(a).
Protons progressively lose energy as they penetrate the sample, with the rate of energy loss proportional to the inverse square of their
velocity following the Bethe formula. This gives the characteristic peak --- known as the Bragg peak --- in damage just before the proton stops.
These Bragg peaks positively skew the dose distribution in our specimens as illustrated in Fig.~\ref{fig:SRIM}(b),
where mode, median and mean doses are indicated.
In Fig.~\ref{fig:schematic} we see CIPs grow slowly with dose, so logarithmic dose axes are appropriate:
then the sub-order-of-magnitude differences between using mode, median, mean, etc.\ are not significant and do not affect our conclusions.
We use the average (mean) dose here.
Each specimen received a dose of $\simeq 6$\,mdpa averaged over its $\simeq 0.2$\,mm thickness
or 7.2\,mdpa averaged over the damaged volume, defined as that which receives local doses greater than 0.1\,mdpa.
This is a reasonable cutoff for CIP formation (cf.\ Fig.~\ref{fig:schematic}) and excludes the undamaged central thickness region.
The average dose rate over the damaged volume is $8.8 \times 10^{-7}$\,dpa\,s$^{-1}$ for the specimen irradiated at 300$^\circ$C
and $7.4 \times 10^{-7}$\,dpa\,s$^{-1}$ for the specimen irradiated at 400$^\circ$C.
These dose rates are $10^4$ times higher than that experienced by the inner wall of a RPV in pressurized water reactor operation.

\subsection{Small-angle neutron scattering (SANS)}
For an introduction to small-angle neutron scattering (SANS) and to magnetic SANS, the reader is referred to Ref.~\cite{Laver12},
and to Refs.~\cite{Mathon97} and \cite{Solt93} for expositions of SANS on RPV model alloys.
Our SANS measurements were performed at the SANS-I instrument at the Swiss Spallation Neutron Source (SINQ).
Neutrons of wavelength $\lambda = 0.6$\,nm with full width at half maximum spread $\Delta \lambda / \lambda = 0.1$ were collimated over distances
between 3\,m and 18\,m before reaching the specimen and detected using a 2D multidetector at distances between 1.7\,m and 20.2\,m behind the specimen,
with distances chosen depending on the desired range of scattering vectors $\mathbf{q}$.
At small scattering angles the magnitude of the scattering vector $q = | \mathbf{q} | = 4 \pi \sin \theta / \lambda$
is approximately proportional to the scattering angle $2 \theta$ measured on the 2D multidetector.
A horizontal electromagnet at the sample position was used to apply a saturating magnetic field of $\mu_0 H = 0.7$\,T perpendicular to the incident neutron beam.
Raw scattering data were corrected for detector noise, neutron absorption within the specimen and air scattering around the specimen.
An unirradiated specimen was also measured (see Fig.~\ref{fig:int}(a) and Sec.~\ref{sec:unirradiated}).

With the specimen in a saturating applied magnetic field $\mathbf{H}$, the magnetic component of scattering exhibits a sine squared dependence
on the angle between $\mathbf{q}$ and $\mathbf{H}$, due to the dipolar nature of the interaction between neutron spin and magnetic moments in the specimen.
This anisotropy in the magnetic scattering allows it to be separated from the scattering from nuclei, which is isotropic.
Magnetic scattering can be comparable or even more intense than nuclear scattering.
For example, Fe has a nuclear bound scattering length of 9.45\,fm~\cite{Dilg74} and a magnetic scattering length of 5.87\,fm,
taking 2.18$\,\mu_\mathrm{B} /$\,ion for bcc iron~\cite{Crangle71}.

\section{Results}
In the following subsections we present our SANS measurements on our proton-irradiated specimens and details of the SANS analysis.
Our results are summarized at the end of this section in Table~\ref{tab:SANS}.

\begin{figure}
\includegraphics[width=8.8cm]{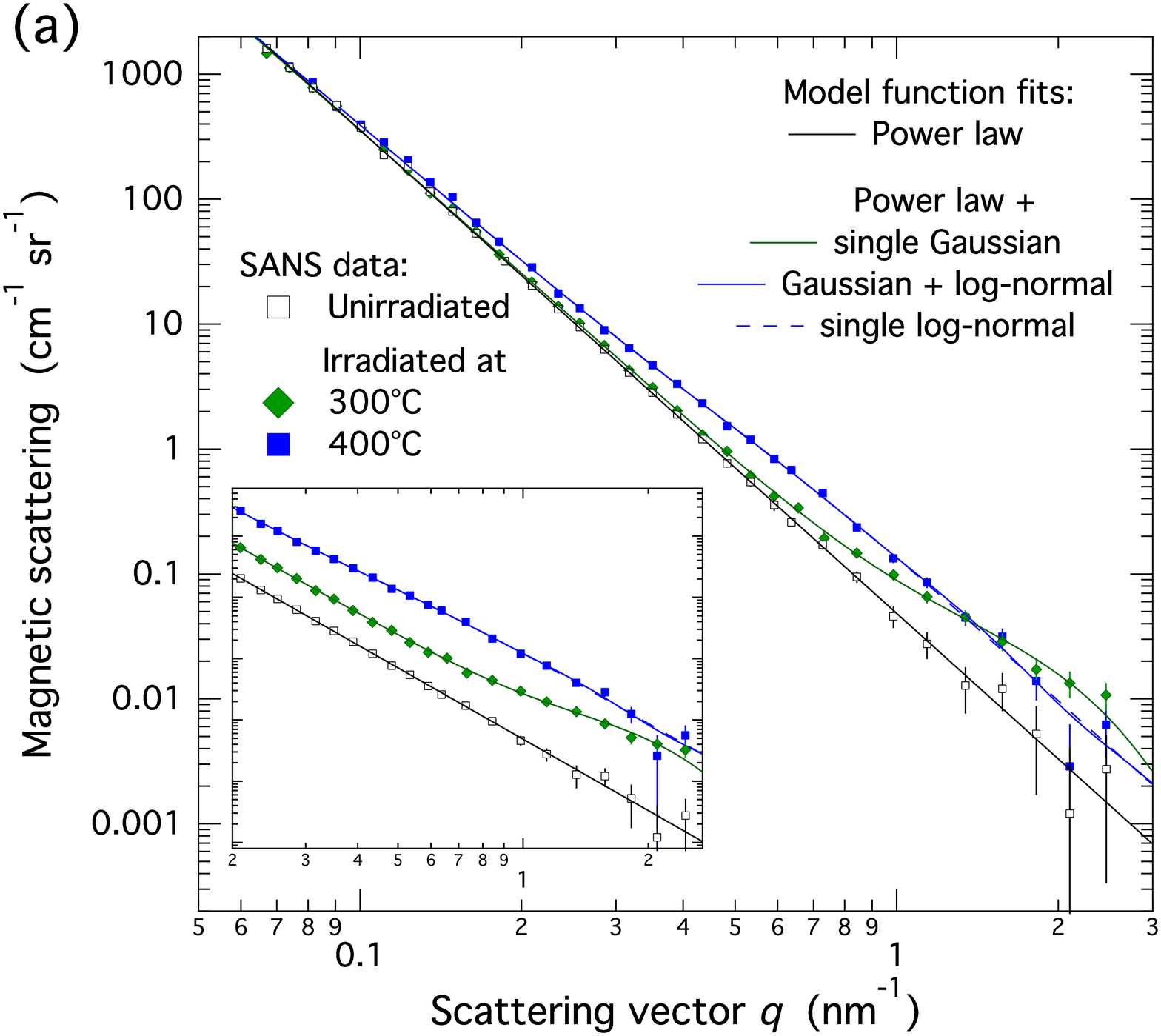}
\includegraphics[width=8.8cm]{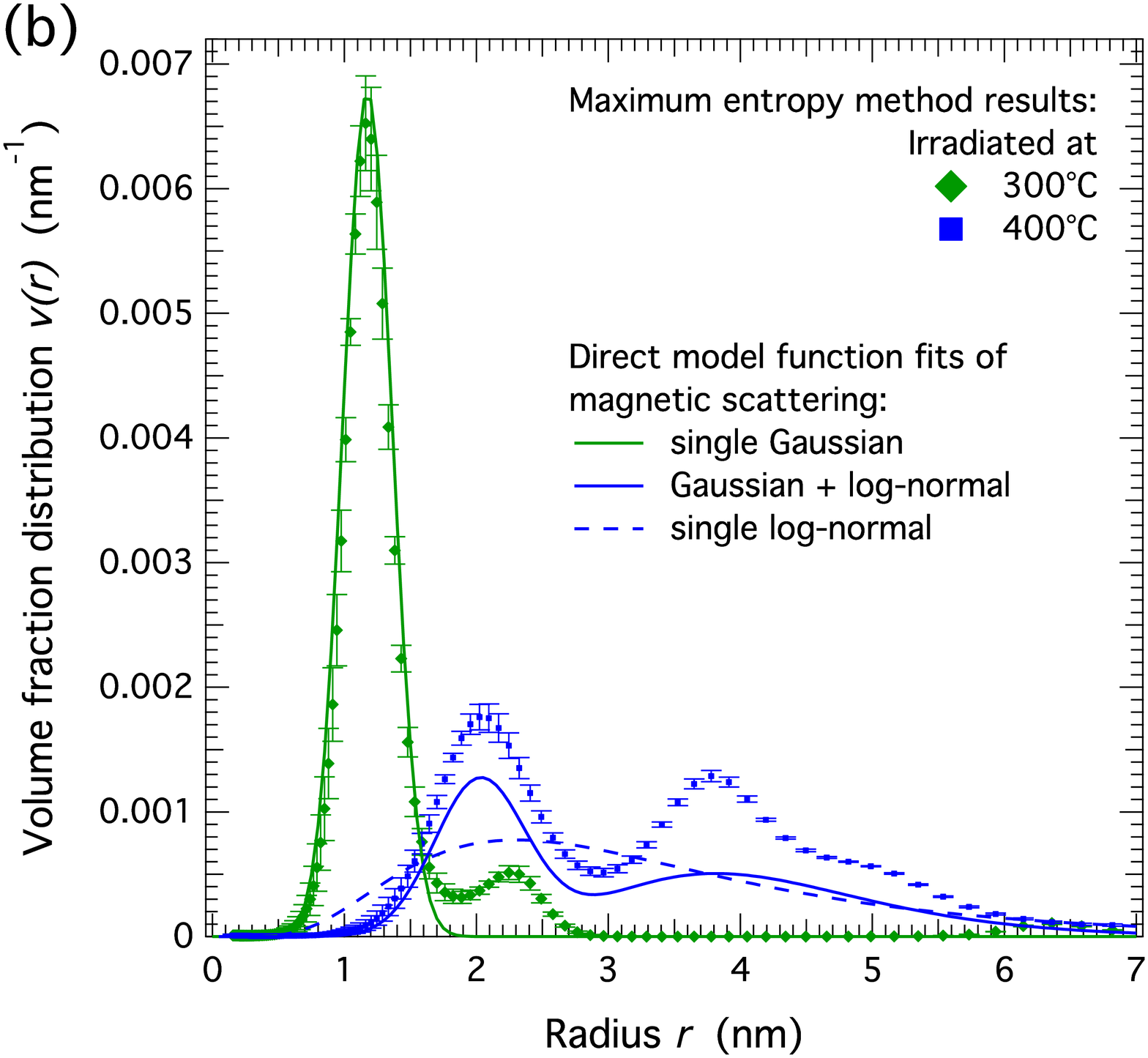}
\caption{\label{fig:int} SANS intensity profiles and fits. (a) Magnetic scattering from our unirradiated and proton irradiated specimens (symbols).
Inset is a zoom on the data at high $q$ with curves offset on the $y$-axis for clarity. Lines show model function fits.
(b) Volume fraction distribution of CIPs reproducing the data in (a) using the maximum entropy method (symbols),
or directly from the model function fits (lines) depicted in (a).
The $v(r)$ scale shown presumes CIPs are non-magnetic.}
\end{figure}

In our setup $\mathbf{H}$ lies perpendicular to the neutron beam and horizontal in the 2D detector plane, so in this direction the scattering is nuclear only.
In the vertical direction the scattering contains both nuclear and magnetic components.
Subtracting the horizontal scattering from the vertical scattering yields the magnetic scattering intensity profiles shown in Fig.~\ref{fig:int}(a).

\subsection{Unirradiated specimen}
\label{sec:unirradiated}
The scattering from the unirradiated specimen (Fig.~\ref{fig:int}(a)) is well described by a power law with exponent $\simeq -4$,
originating from scatterers at much larger scales than those probed by our measured $q$-range.
This can be understood as follows:
in the limit of dilute similar scatterers, the scattered intensity is proportional to the square of the form factor of an individual scatterer.
For example, the form factor for a uniform sphere of radius $R$ is $F(q R) = 3 j_1(q R) / q R$, where the spherical Bessel function $j_1(x) = (\sin x - x \cos x) / x^2$.
At large $q$, that is at $q D \gg 1$, where $D$ is the diameter of the scatterer, form factors typically decay as a power law in $q$.
The power law observed from the unirradiated specimen (Fig.~\ref{fig:int}(a)) extends to the minimum $q$ of $0.06$\,nm$^{-1}$ probed in these measurements
and hence the scatterers causing it must have diameters $\gg 17$\,nm.
Given the significant carbon content (Table~\ref{tab:composition}) and bainitic microstructure of our specimens, these scatterers are expected to be carbides.
As these features are much larger than the CIPs expected under irradiation, we do not focus further attention on this power-law scattering component.

\subsection{Maximum entropy analysis}
In the irradiated specimens, in addition to the power law there is a clear bump in the scattering profile at $\approx 2$\,nm$^{-1}$ for the 300$^\circ$C irradiated specimen
and a broader feature between $\approx 0.4$\,nm$^{-1}$ and $\approx 2$\,nm$^{-1}$ for the 400$^\circ$C irradiated specimen (Fig.~\ref{fig:int}(a)).
Seeking an unprejudiced insight into these irradiation-caused features,
we subtract the unirradiated scattering profile from each irradiated scattering profile and apply the maximum entropy method to the results.
In this method~\cite{Potton88,Jemian91}, a maximum entropy algorithm is used to fit a size distribution $v(r)$ of uniform spherical scatterers to the scattering profile $I(q)$ via
\begin{equation} I(q) = |\Delta \rho|^2 \int_0^{\infty} \, \tfrac43 \pi r^3 \, F^2(qr) \, v(r) \, \mathrm{d}r \label{eq:int} \end{equation}
If the difference in scattering length density $\Delta \rho$ between scatterer and matrix is known,
then the $v(r)$ fitted in Eq.~\eqref{eq:int} is the {\em volume fraction distribution} i.e.\ integrating $v(r)$ over all $r$
yields the volume fraction $V$ occupied by scatterers. The scattered intensity scales as $V \, |\Delta \rho|^2$
and we provide this quantity in case future work should yield improved knowledge of $\Delta \rho$.
When quoting volume fractions we assume the scatterers are non-magnetic and reside in a ferromagnetic bcc Fe matrix i.e.\ that
the magnetic scattering contrast $|\Delta \rho|^2 = 2.48 \times 10^{-7}$\,nm$^{-4}$. We argue that this assumption is valid in the discussion (Sec.~\ref{sec:CIPmag}).
Our quoted volume fractions also regard CIPs as forming in the damaged volume of irradiated specimens and not in their undamaged central regions (Fig.~\ref{fig:SRIM}).
Throughout this article we weight averages of the size distribution, such as mean radius, by number density
so that these averages may be directly compared with atom probe tomography results in the literature.
The number density distribution is readily derived from $v(r) = \frac43 \pi r^3 \, n(r)$.

The $v(r)$ found by applying the maximum entropy method to the irradiated minus unirradiated scattering profiles are shown by symbols in Fig.~\ref{fig:int}(b).
Uncertainties in $v(r)$ are estimated by running the maximum entropy code with different initial conditions and taking the standard deviation of the results.
The $v(r)$ for the 300$^\circ$C irradiated specimen (green diamonds in Fig.~\ref{fig:int}(b)) shows two peaks that both fit well to Gaussian sphere size distributions.
The taller peak has mean radius 1.1\,nm and standard deviation 0.2\,nm.
Gaussian distributions are often found to fit precipitates in Cu-containing Fe alloys (see e.g.\ \cite{Mathon97, Dohi10}).
For the 400$^\circ$C irradiated specimen the maximum entropy analysis returns a bimodal size distribution (blue squares in Fig.~\ref{fig:int}(b)),
with a long tail at high $r$ and perhaps even a third peak at $r \approx 5$\,nm. The $v(r)$ fits well to the sum of two spherical size distributions:
a Gaussian sphere distribution of mean radius 1.8\,nm and standard deviation 0.4\,nm 
plus a log-normal sphere distribution of median radius 3.4\,nm and width $\sigma = 0.25$,
where $\sigma$ is the standard deviation of $\ln r$.

\subsection{Direct model fitting}
It proves instructive to check whether the maximum entropy method over-fits the scattering data.
We take the previous spherical size distribution functions fitted to the $v(r)$ output (symbols in Fig.~\ref{fig:int}(b))
and see if their corresponding scattering model functions directly fit our SANS scattering profiles (Fig.~\ref{fig:int}(a)),
including a power-law function to account for the microstructural background at larger scales.

For the 300$^\circ$C irradiated specimen we find the small peak at larger radius is not supported by direct fitting,
as negative volume fractions then result for this feature.
The large peak is confirmed: a single Gaussian-distributed sphere scattering model function plus a power law fit is shown by the solid green line in Fig.~\ref{fig:int}(a)
and the corresponding $v(r)$ distribution function is plotted in Fig.~\ref{fig:int}(b), also with a solid green line.
The height of this directly-fitted distribution agrees remarkably well with the maximum entropy result
(green diamonds in Fig.~\ref{fig:int}(b)), yielding a volume fraction $V$ of 0.32(3)\%.
We note that attempting to fit a log-normal distributed sphere scattering model function plus a power law directly to the $I(q)$ produces a 45\% increase in $\chi^2$
compared to the Gaussian-distributed model, so the 300$^\circ$C data evidently do not support a long tail in the size distribution.

For the 400$^\circ$C irradiated specimen, a positive skew in the size distribution is suggested by maximum entropy fitting
and reinforced by direct fitting of $I(q)$.
However the pronounced bimodal nature suggested by the maximum entropy method is less apparent in direct fits.
The volume fractions from direct fitting of Gaussian plus log-normal distributed sphere scattering model functions including a power law (solid blue lines in Fig.~\ref{fig:int}(a) and (b)) are
0.11(2)\% and 0.13(3)\% for the Gaussian and log-normal components respectively.
These are smaller than those suggested by the maximum entropy analysis, as can be seen in Fig.~\ref{fig:int}(b)
by comparing the directly-fitted scattering model distribution (solid blue line) with the maximum entropy result (blue squares).
This difference illustrates the need for caution when using maximum entropy methods or other numerical refinements with very many degrees of freedom,
since extraneous fluctuations in $I(q)$ may readily be over-fitted.
Furthermore, we are not fully convinced that a multi-modal distribution can be discerned in our 400$^\circ$C data.
For example, directly fitting a single log-normal distributed sphere scattering model function plus power law to $I(q)$ (dashed blue line in Fig.~\ref{fig:int}(a)) increases $\chi^2$ by only 2\%.
This unimodal size distribution, overplotted as the dashed blue line on Fig.~\ref{fig:int}(b) for comparison, has a median radius of 1.3\,nm and width $\sigma = 0.53$.

\begin{table*}
\caption{\label{tab:SANS} Summary of SANS results on proton irradiated specimens.}
\centerline{
\begin{tabular}{llcccccccc}
\hline
                 &                 &       $M$, Magnetic      &       $N$, Nuclear       & $M/N$ & $V$, Volume &         Number        &  Mean  &  Standard \\
$T_\mathrm{irr}$ &                 & $V \, |\Delta \rho|^2$ & $V \, |\Delta \rho|^2$ & ratio &    fraction   &         density       & radius & deviation \\
  ($^\circ$C)    & Analysis method & ($10^{-10}$\,nm$^{-4}$)  & ($10^{-10}$\,nm$^{-4}$)  &       &      (\%)     & ($10^{22}$\,m$^{-3}$) &  (nm)  &    (nm)   \\ \hline
300 & Single Gaussian sphere $I(q)$ fit   & 7.9(8) & 2.6(1) & 3.0(14) & 0.32(3) & 55  & 1.1 & 0.2 \\
    & Azimuthal sine-squared + const.\ fit &        &        &  2.4(5) &         &     &     &     \\ \hline
400 & Two-component $I(q)$ fit:           &        &        &         &         &     &     &     \\
    & $\quad$ Gaussian sphere component   & 2.7(6) & 0.1(5) &  19(62) & 0.11(2) & 3.8 & 1.8 & 0.4 \\
    & $\quad$ log-normal sphere component & 3.1(7) & 1.2(3) &  2.5(8) & 0.13(3) & 0.6 & 3.5 & 0.9 \\
    & Averaged over both components       & 5.8(9) & 1.4(5) & 4.2(18) & 0.23(4) & 4.4 & 2.1 & 0.7 \\
    & Single log-normal $I(q)$ fit        & 6.7(9) & 1.8(3) &  3.7(8) & 0.27(4) & 8.0 & 1.5 & 0.9 \\ \hline
\end{tabular}
}
\end{table*}

\subsection{$M/N$ ratios}
The nuclear scattering $I(q)$ profiles are fitted directly with the same scattering model functions as the magnetic scattering profiles,
including a constant term to account for incoherent scattering.
The fitted constant terms, 0.007(1)\,cm$^{-1}$\,sr$^{-1}$ for the 400$^\circ$C irradiated specimen
and 0.006(2)\,cm$^{-1}$\,sr$^{-1}$ for the 300$^\circ$C irradiated specimen,
are consistent with expectations e.g.\ the calculated incoherent scattering for a randomly disordered alloy of our composition (Table~\ref{tab:composition}) is 0.0062\,cm$^{-1}$\,sr$^{-1}$.
The fitted nuclear scattering intensities $N$ are listed in Table~\ref{tab:SANS} along with the magnetic scattering intensities $M$
and the dimensionless ratio $M/N$, which provides information as to the chemical composition of scatterers.
For the 400$^\circ$C irradiated specimen, the smaller Gaussian sphere component has barely discernible nuclear scattering,
resulting in a high $M/N$ ratio with large uncertainty.
Given that we are not convinced that these data support two size features, we 
also list in Table~\ref{tab:SANS} the single log-normal sphere model fit results (blue dashed lines in Fig.~\ref{fig:int}),
as well as the sum of the two-component fits. The latter yields a combined  $M/N$ ratio of $4.2 \pm 1.8$
and a combined size distribution of mean radius 2.1\,nm and standard deviation 0.7\,nm.

\begin{figure}
\begin{center}
\includegraphics[width=7.6cm]{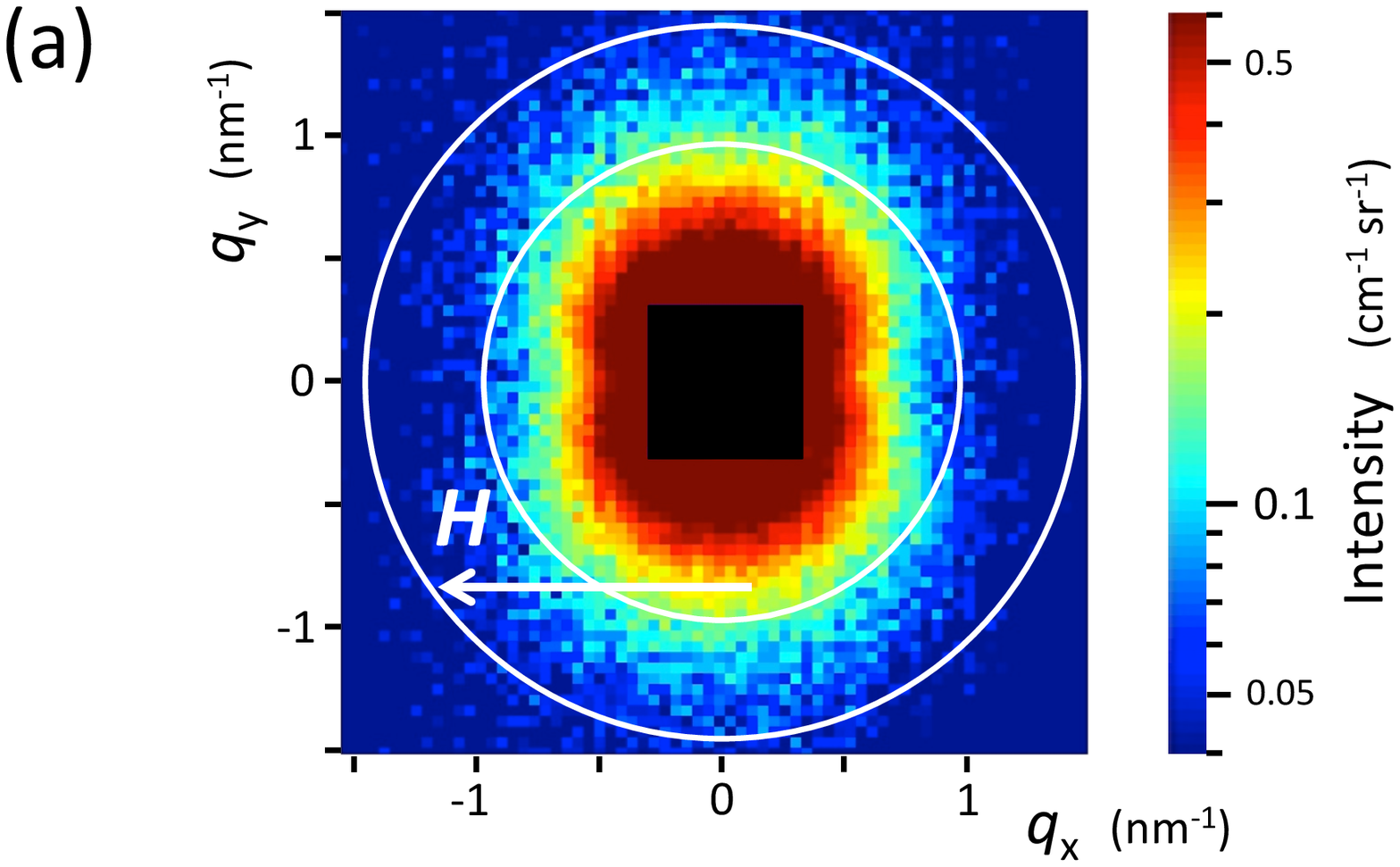} \hspace{0.5cm} % space added for single-column version
\includegraphics[width=8.1cm]{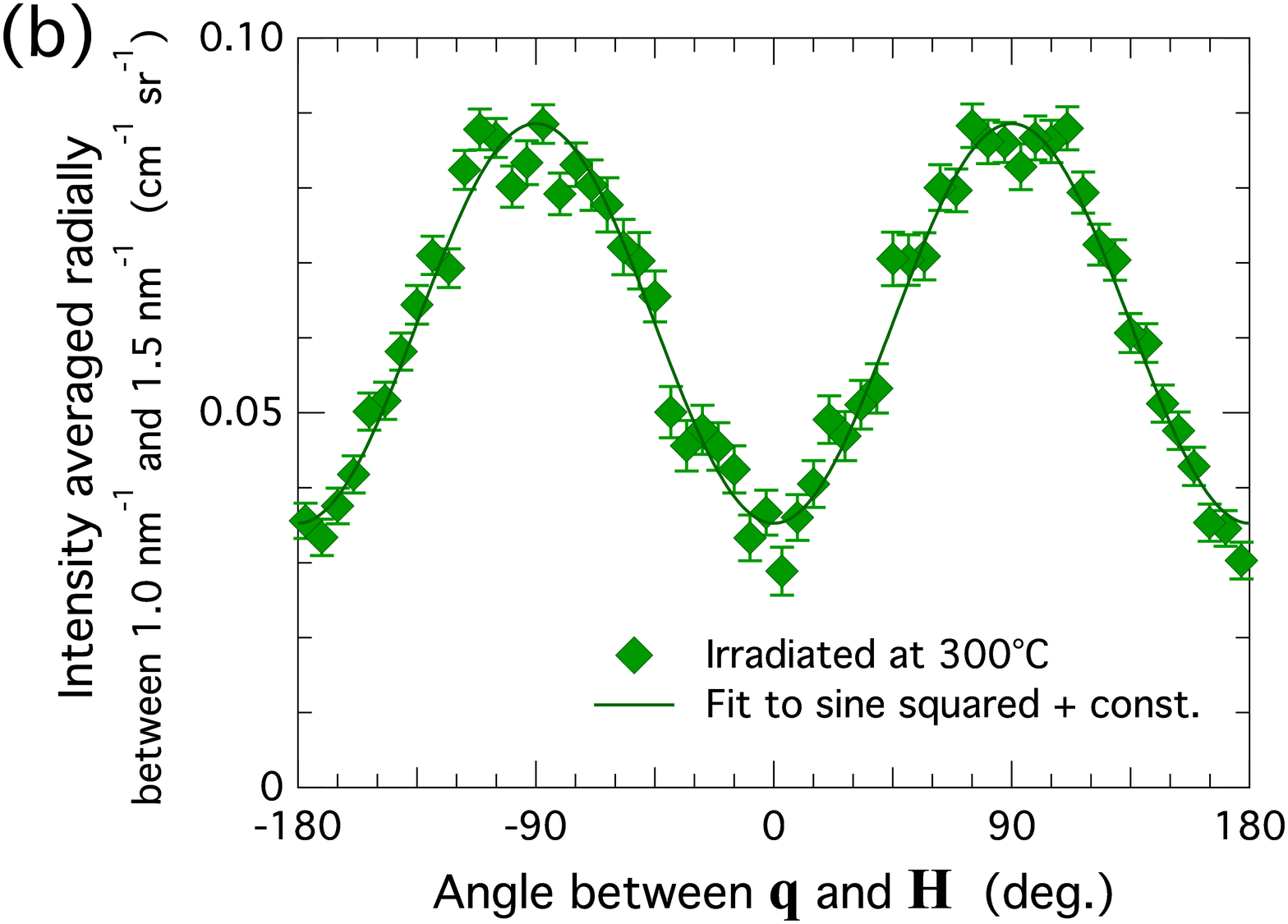}
\end{center}
\caption{\label{fig:azi} SANS from 300$^\circ$C proton-irradiated specimen. (a) 2D image of SANS multidetector. The unscattered neutron beam
at the image centre is blocked by a beamstop. The applied magnetic field $\mathbf{H}$ lies horizontal and perpendicular to the neutron beam.
(b) Azimuthal cut, averaging radially over the $1.0 \leq q \leq 1.5$\,nm$^{-1}$ range in which scattering from CIPs is most apparent, depicted by the white annulus in (a).
The line is a fit with a sine-squared plus constant term function describing magnetic and nuclear scattering respectively.}
\end{figure}

The 300$^\circ$C irradiated specimen has a $M/N$ ratio of $3.0 \pm 1.4$ from the $I(q)$ analysis.
Here the narrow size distribution gives well-defined features in $I(q)$ and we can improve the uncertainty on $M/N$ by
adding more scattering signal to the analysis (beyond vertical and horizontal sectors) and examining the scattering anisotropy on the 2D SANS detector (Fig.~\ref{fig:azi}(a)).
The sine-squared anisotropy, due to the dipolar interaction between neutron spin and magnetic moments in the specimen, is clear in Fig.~\ref{fig:azi}(a).
In Fig.~\ref{fig:azi}(b) we show an azimuthal cut over the $1.0 \leq q \leq 1.5$\,nm$^{-1}$ range;
this is fitted to a sine-squared plus constant term function: the amplitude of the sine-squared component is the magnetic intensity and the constant is the nuclear intensity.
After correcting for the background power-law contribution, we obtain a $M/N$ ratio of $2.4 \pm 0.5$.

\section{Discussion}
We begin our discussion by comparing our $M/N$ ratio measured by SANS with previous atom probe tomography studies in Sec.~\ref{sec:CIPcomp}.
This comparison allows us to deduce likely compositions for CIPs in our proton irradiated specimens (Fig.~\ref{fig:ternary}).
It also aids, in a self-consistent manner, justification of some assumptions made when calculating theoretical $M/N$ ratios (Sec.~\ref{sec:CIPmag}).
In Sec.~\ref{sec:sizedistributionstart} we compare the volume fraction and radius of CIPs in our 300$^\circ$C proton-irradiated specimen
with literature on RPV steels with similar copper and high nickel content. This literature data set consists of neutron irradiated and Fe ion irradiated steels
spanning over 6 orders of magnitude in dose rate, from neutron dose rates $\approx 0.1$\,ndpa\,s$^{-1}$ matching RPV operational conditions~\cite{Bergner08,Hyde01,Eason07},
to Fe ion dose rates up to 150\,000\,ndpa\,s$^{-1}$~\cite{Almirall20}.
The results from our 300$^\circ$C proton irradiation at 880\,ndpa\,s$^{-1}$ dose rate are found to be in good quantitative agreement
with the literature data set, bearing in mind the known effects of nickel (Sec.~\ref{sec:nickel}) and its high content (3.3\,at.\%) in our specimen.
We compare methods to normalize for dose rate effects and find
power-law scaling of dose works well for CIP volume fractions (Sec.~\ref{sec:volfracscaling} and Fig.~\ref{fig:scaling}), but not for CIP sizes (Sec.~\ref{sec:radius} and Fig.~\ref{fig:radscaling}).
Carbon content is identified to be at least as significant as dose rate in determining CIP sizes (Fig.~\ref{fig:radfactors}(c));
we conclude these two factors work cooperatively in Sec.~\ref{sec:carbon}.

\subsection{CIP solute composition}
\label{sec:CIPcomp}
Compositions reported by atom probe tomography (APT) studies can be directly compared to SANS measurements of $M/N$ ratio, or `A' ratio $= M/N + 1$.
The numerator and denominator in $M/N$ depend respectively on the magnetic and nuclear contrasts i.e.\ on the squares of differences in scattering length densities between CIP and matrix.
Assuming the atomic volume within the CIP is the same as that in the matrix, $M/N$ reduces to an expression in terms of scattering lengths
\begin{equation} \frac{M}{N} = \left( \frac{ 5.87 }{ 9.45 - \left\{ 7.72 a_\mathrm{Cu} + 10.3 a_\mathrm{Ni} - 3.75 a_\mathrm{Mn} + 4.15 a_\mathrm{Si} \right\} } \right)^2 \label{eq:MN} \end{equation}
where it is additionally assumed that CIPs are non-magnetic. These assumptions will be justified in Sec.~\ref{sec:CIPmag}.
$a_j$ is the atomic fraction of element $j$ in a CIP {\em excluding} Fe. Decimal values are elemental neutron scattering lengths in fm.
Terms in minor enriching elements such as Mo and Cr are omitted since their effects are negligible here.

We begin by comparing the results of Zelenty {\it et al.}~\cite{Zelenty16} on precisely the model steels examined in our study.
Following thermal ageing at 405$^\circ$C for 16 months, Zelenty {\it et al.}~\cite{Zelenty16} report precipitates throughout the ferrite matrix and also in the cementite phase.
The average composition from APT of precipitates within cementite is reported to be 12.78\,at.\% Cu, 43.48\,at.\% Ni, 24.67\,at.\% Mn, 16.96\,at.\% Si, 1.56\,at.\% Mo and 0.55\,at.\% Cr,
ignoring Fe and C~\cite{Zelenty16}.
Using Eq.~\eqref{eq:MN} this gives a $M/N$ ratio of 2.06.
An APT reconstruction of one precipitate (Fig.~1 of Ref.~\cite{Zelenty16}) appears to show Ni- and Si-rich regions forming as appendages to a Cu-rich core,
consistent with the growth of appendages rich in Ni, Mn and Si (NMS) observed by APT in other RPV steels~\cite{Wells14,Shu18a,Odette19,Almirall20}.
Ignoring other elements, the relative NMS composition is 51.1\,at.\% Ni, 29.0\,at.\% Mn and 19.9\,at.\% Si.
In terms of Si content this lies midway between the $G$ (Ni$_{16}$Mn$_6$Si$_7$) and the $\Gamma_2$ (Ni$_3$Mn$_2$Si)
phases proposed as culminating thermodynamic endpoints~\cite{Odette19}
but is closer in Ni and Mn content to $\Gamma_2$, congruous with APT~\cite{Wells14} and X-ray diffraction~\cite{Sprouster16} results
indicating that $\Gamma_2$ rather than $G$ forms in high Ni (1.7\,at.\%) RPV steel.
For our 400$^\circ$C proton-irradiated specimen, our two-feature fits to SANS $I(q)$ profiles suggest large $\simeq 7$\,nm diameter scatterers with $M/N = 2.5 \pm 0.8$
and small $3.6$\,nm diameter scatterers with low nuclear contrast and high $M/N$ ratio (Table~\ref{tab:SANS}).
Though we are not convinced that two size features are supported by our data, this result may be reconciled with a picture of small copper cores ($M/N$ ratio $= 11.5$)
surrounded by large NMS appendages ($M/N$ ratio $=1.74$ for Ni:Mn:Si $=$ 51:29:20\,at.\%) growing in random directions from the cores.

\begin{figure*}[h]
\begin{center}
\includegraphics[width=16.5cm]{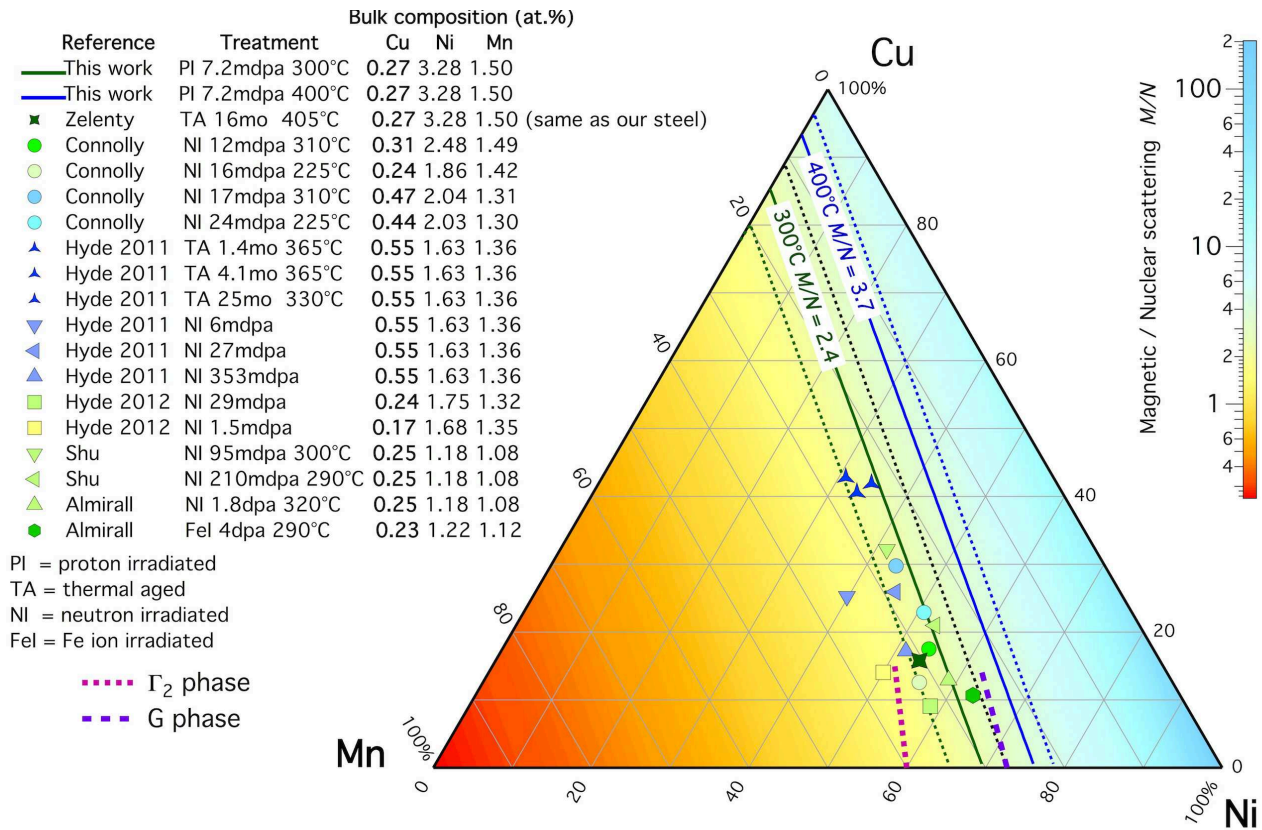}
\end{center}
\caption{\label{fig:ternary}Ternary plot showing relative Cu, Ni and Mn compositions (in at.\%) of CIPs consistent
with our SANS data on proton irradiated specimens (solid contours; dotted contours are standard uncertainties),
compared to APT results on high Ni, Cu-containing steels from the literature (symbols).
The dashed oval depicts the likely CIP composition in our 300$^\circ$C proton-irradiated specimen explained in the text (Sec.~\ref{sec:CIPcomp}).
References are Zelenty {\it et al.}~\cite{Zelenty16}, Connolly {\it et al.}~\cite{Connolly15}, Hyde, Sha {\it et al.}~\cite{Hyde11},
Hyde, Boothby {\it et al.}~\cite{Hyde12}, Shu {\it et al.}~\cite{Shu18a} and Almirall {\it et al.}~\cite{Almirall20}.
Calculated SANS $M/N$ ratios (background colour and contours) include 20\,at.\% Si in the CIP composition.
}
\end{figure*}
The term in curly braces $\{ \ldots \}$ in Eq.~\eqref{eq:MN} is the average nuclear CIP scattering length.
For the precipitates reported by Zelenty {\it et al.}~\cite{Zelenty16} it evaluates to 5.36\,fm.
As this is close to the nuclear scattering length of Si (4.15\,fm), our results are anticipated to be fairly insensitive to the relative Si content of CIPs
e.g.\ an increase of 5\,at.\% in Si content suppresses $M/N$ from 2.06 to 1.98.
The same suppression is achieved by an increase in relative Mn CIP content of just 0.6\,at.\%, demonstrating that $M/N$ is sensitive to Mn.
It is also sensitive to Cu and Ni but not to these elements separately, as illustrated by the ternary Cu-Ni-Mn diagram (Fig.~\ref{fig:ternary})
where lines of constant $M/N$ are approaching parallel to the Cu axis.
The $M/N$ values displayed in Fig.~\ref{fig:ternary} are calculated using Eq.~\ref{eq:MN} with a fixed CIP Si content of 20\,at.\%,
appropriate for the composition reported by Zelenty {\it et al.}~\cite{Zelenty16}.

In Fig.~\ref{fig:ternary} we also overplot CIP compositions reported by other APT studies on high Ni, Cu-containing steels~\cite{Connolly15,Hyde11,Hyde12,Shu18a,Almirall20}.
First we note our measured $M/N = 2.4(5)$ for the 300$^\circ$C proton-irradiated specimen is broadly consistent with the ensemble of CIP compositions from the APT literature.
Looking more closely at the latter, we see the relative Ni content of CIPs tends to increase with neutron fluence at the expense of Cu
(cf.\ triangle symbols in Fig.~\ref{fig:ternary}~\cite{Hyde11,Shu18a,Almirall20}),
consistent with NMS enrichment around fast-forming Cu cores.
Thermally aged compositions have somewhat higher Cu content compared to neutron irradiated specimens,
as noted by Hyde, Sha {\it et al.}~\cite{Hyde11} for RPV weld steels and illustrated in Fig.~\ref{fig:ternary} (three-pointed stars and blue triangles)
by their results on steels of 0.55\,at.\% Cu and 1.63\,at.\% Ni content.
At first glance this disparity might be ascribed to thermally aged CIPs being in an earlier stage of growth.
However, high relative Cu contents ($\gtrsim 30$\,at.\% averaged over the CIP volume) are also observed in CIPs that lie demonstrably in a coarsening stage,
from their decreasing number density and plateauing hardness, after long-time ageing for 10 years at 365$^\circ$C or at 330$^\circ$C
in steels that contain less Cu (0.44\,at.\%) and similar Ni (1.66\,at.\%)~\cite{Styman18}.
We suggest the altogether greater NMS content of CIPs after neutron irradiation compared to thermal ageing
implies that the factor by which diffusion is enhanced by radiation is larger for Ni than it is for Cu.
This is possible e.g.\ when different diffusion mechanisms are active for each solute.
The diffusion of copper is dominated by the vacancy mechanism~\cite{Vincent05, Barashev06, Messina14}
but dumbbell-assisted diffusion is operative as well for nickel~\cite{Messina20}.

The higher bulk Ni to Cu ratio explains the higher Ni and lower Cu content of the precipitates in cementite reported by Zelenty {\it et al.}~\cite{Zelenty16} (four-pointed star in Fig.~\ref{fig:ternary})
compared to the thermally aged steels of Hyde, Sha {\it et al.}~\cite{Hyde11} (three-pointed stars in Fig.~\ref{fig:ternary}) and of Ref.~\cite{Styman18}.
Given that solute diffusion coefficients in cementite may be markedly different to those in ferrite, it is remarkable that the CIP composition reported by Zelenty {\it et al.}~\cite{Zelenty16}
lies so close to the compositions reported by Connolly {\it et al.}~\cite{Connolly15} in high Ni, neutron-irradiated steels of similar Cu content (green circles in Fig.~\ref{fig:ternary}).
This coincidence is possibly due to the balance of two opposing effects:
irradiation of the latter steels rather than thermal ageing of the former, favouring less Cu and more Ni in CIPs, counterbalancing
the lower bulk Ni to Cu ratio of the latter steels, favouring more Cu and less Ni in CIPs.
With this in mind, we can tentatively estimate the composition of CIPs in our 300$^\circ$C proton-irradiated steel;
this estimate is sketched by the dashed oval in Fig.~\ref{fig:ternary}.
The uncertainty and value of the $M/N$ ratio measured by SANS constrain the short axis of this oval and its lateral position on the ternary diagram (Fig.~\ref{fig:ternary}),
while its position along the $M/N$ contour is an estimate accounting for the irradiation versus thermal-ageing effect and that our proton-irradiated steel
has exactly the same bulk composition as the thermally aged steel of Zelenty {\it et al.}~\cite{Zelenty16}, whose CIP composition was measured by APT (four-pointed star in Fig.~\ref{fig:ternary}).
The centre of this oval corresponds to a relative Cu-Ni-Mn-Si CIP composition of $\approx 11$\,at.\% Cu, $\approx 47$\,at.\% Ni, $\approx 22$\,at.\% Mn and $\approx 20$\,at.\% Si.

The higher proton irradiation temperature of 400$^\circ$C clearly produces CIPs containing on average more Cu or Ni or both, as shown by our single-feature fit results
(blue contours in Fig.~\ref{fig:ternary}).
If we accept that our double-feature fit to these data corresponds to Cu cores surrounded by randomly orientated NMS appendages,
then this picture favours greater Cu enrichment, since the smaller feature corresponding to pure Cu cores
constitutes nearly half the precipitate volume fraction (cf.\ Table~\ref{tab:SANS}).

\subsection{CIP magnetism, vacancy and Fe content}
\label{sec:CIPmag}
Here we justify the assumptions made in the expression for $M/N$ of Eq.~\eqref{eq:MN},
starting with the assumption that the atomic volume within CIPs $\Omega_\mathrm{p}$ is the same as that of the matrix $\Omega_\mathrm{m}$.
Any difference between $\Omega_\mathrm{m} = 11.78 \times 10^{-30}$\,m$^3$ and $\Omega_\mathrm{p}$
can be included in the $M/N$ calculation simply by multiplying the average nuclear CIP scattering length (the $\{ \ldots \}$ term in Eq.~\eqref{eq:MN})
by $\delta = \Omega_\mathrm{m} / \Omega_\mathrm{p}$, under the assumption that CIPs are non-magnetic.
Calculated values for $\delta$, excluding vacancies, of eventual crystallographic CIP phases are close to unity, e.g.\ $\delta = 1.000$ for fcc Cu,
$\delta = 0.986$ for $G$ phase Ni$_{16}$Mn$_6$Si$_7$~\cite{Ahmed18}, and their inclusion would not have a significant effect on our results.
An explicit vacancy fraction $z_\mathrm{v}$ in CIPs can similarly be included into Eq.~\eqref{eq:MN} by multiplying the average nuclear CIP scattering length by $(1-z_\mathrm{v})$,
where CIPs are again assumed to remain non-magnetic. For voids, $z_\mathrm{v} = 1$ yields $M/N = 0.4$.
We see the effect of increasing $z_\mathrm{v}$ is to suppress $M/N$
and so to shift the calculated $M/N$ values, that is the background colour and contours representing our SANS results in Fig.~\ref{fig:ternary},
away from the Mn corner and towards the Ni-Cu edge of the ternary diagram.
The ensemble of CIP compositions from the APT literature no longer overlap the uncertainties of our measured SANS $M/N$ when $z_\mathrm{v} \gtrsim 0.1$.

Including a ferromagnetic CIP moment into the $M/N$ calculation dilutes the magnetic contrast, producing decreases of calculated $M/N$ ratios across the ternary diagram.
This shifts the calculated $M/N$ contours representing our SANS results in Fig.~\ref{fig:ternary}
to the right of the ternary diagram, away from the ensemble of CIP compositions from the APT literature.
These latter compositions no longer overlap within the uncertainties of our SANS $M/N$ for CIP moments of $\gtrsim 0.2$\,$\mu_\mathrm{B} /$\,ion.
There are also good conjectural reasons for small or zero CIP magnetic moments.
Copper is observed to carry a negligible magnetic moment (0.002(94)\,$\mu_\mathrm{B} /$\,ion) from polarized neutron diffuse scattering measurements on
a bcc Fe$_{96.1}$Cu$_{3.9}$ alloy~\cite{Kajzar79}.
In binary bcc Fe$_{1-x}$Cu$_x$ alloys the measured magnetization change with Cu concentration $x$ is linear for $x \lesssim 20$\,at.\%
and lies very close to (but not exactly at) the line given by simple dilution,
where solute atoms have zero magnetic moment and no effect on Fe moments~\cite{Aldred68,Sumiyama84,Ma93}.
This apparent spin decoupling between solute and Fe electrons is unusual. It is due to Cu and Fe $d$ states being well separated in energy~\cite{Chien86},
as substantiated by first principles calculations that find solitary Cu ions in bcc Fe carry
a small magnetic moment $\simeq 0.1$\,$\mu_\mathrm{B}$~\cite{Drittler89,Akai85,Leonard82,Kontsevoi95,Razee97,Rahman10}.
They also enhance the moment of nearest neighbour Fe atoms by $\simeq 0.2$\,$\mu_\mathrm{B}$ but do not appreciably affect farther Fe moments~\cite{Kontsevoi95,Razee97,Rahman10}.
These findings explain why the magnetization decays slightly more slowly with $x$ than expected from simple dilution~\cite{Aldred68}.
Including 0.1\,$\mu_\mathrm{B} /$ Cu ion reduces the calculated $M/N$ ratio for pure copper precipitates from 11.5 to 10.5;
this would affect our interpretation of the double-feature fit to our 400$^\circ$C irradiated data only by increasing the inferred volume fraction $V$ of the small features
(ostensibly Cu cores) by $\simeq 10$\,\%.
For our 300$^\circ$C irradiated results, the modest Cu content $\approx 11$\,at.\% anticipated in CIPs
means that any corrections due to small $\approx 0.1$\,$\mu_\mathrm{B}$ Cu moments are insignificant.

In disordered alloys Mn-Mn interactions tend to be antiferromagnetic and suppress the surrounding magnetism,
as observed in fcc Ni$_{1-y}$Mn$_y$ alloys for example, where the ferromagnetism of pure nickel collapses at $y \approx 10$\,at.\%~\cite{Podolak09}.
If the NMS-part of CIPs should order crystallographically, we might likewise expect this phase to be non-magnetic in our SANS measurements.
Bulk $G$ phase (Ni$_{16}$Mn$_6$Si$_7$) for instance orders antiferromagnetically at $T_\mathrm{N} \simeq 197$\,K
with Mn carrying all the magnetic moment~\cite{Ahmed18}.
At room temperature it is paramagnetic with a small susceptibility~\cite{Ahmed18} that would yield 0.25\,$\mu_\mathrm{B}$ per formula unit
$= 0.009$\,$\mu_\mathrm{B} /$\,ion in the 0.7\,T applied field of our SANS measurements i.e.\ a negligible moment.

Having argued that the magnetic moment of CIPs in our specimens is insignificant based on theoretical grounds
as well as for consistency with APT-measured compositions from the literature, we note an unfortunate corollary
is that our SANS results are insensitive to the iron content of CIPs.
This can be readily understood from Eq.~\eqref{eq:MN} where we see that if we attempt to include a fractional atomic Fe content $z_\mathrm{Fe}$ in CIPs,
while preserving the relative solute compositions
and assuming simple magnetic dilution, then a factor $(1-z_\mathrm{Fe})$ enters before each of the terms in Eq.~\eqref{eq:MN} and so divides out.
To have any appreciable effect on the $M/N$ ratio, the average magnetism of CIPs should be strongly non-linear in $z_\mathrm{Fe}$,
for which there is presently no quantifiable justification as far as we are aware.
The reducing scattering contrast due to increasing $z_\mathrm{Fe}$ does mean that the inferred CIP volume fraction $V$ increases as $1/(1-z_\mathrm{Fe})^2$.
This can be used to set an upper limit on the possible $z_\mathrm{Fe}$, corresponding to depletion from the matrix of any one CIP-enriching solute~\cite{Miller03}.
This argument is not useful here since CIPs lie in an early stage of growth for our proton irradiation doses, far from matrix depletion.
However we shall see shortly in Sec.~\ref{sec:nickel} and in Sec.~\ref{sec:volfracscaling}
that our CIP volume fractions inferred with $z_\mathrm{Fe}=0$ are consistent with other measurements from the literature.

\subsection{CIP size distribution}
\label{sec:sizedistributionstart}
\subsubsection{Known effects of nickel and manganese}
\label{sec:nickel}
The volume fraction and average size of CIPs in our $300^\circ$C proton-irradiated specimen are in broad agreement with reports on RPV steels irradiated to similar doses
with high Ni and similarly high Cu content~\cite{Glade06,Eason07}. At 1.1\,nm mean radius, our CIPs are approximately half the size of copper precipitates
forming under irradiation in binary Fe-Cu alloys~\cite{Mathon97,Meslin10}.
This is due to the presence of nickel and manganese, which are known to segregate at the CIP-matrix interface
and lead to more numerous but smaller CIPs with increasing Ni~\cite{Monzen04,Miller07} or Mn~\cite{Miller03,Glade05,Meslin10,Shu18m,Shu18p} content,
signifying a reduction in interfacial energy~\cite{Isheim06,Monzen04,Styman15,Styman18}
and/or a lowering of the work needed to form a critical CIP nucleus~\cite{Zhang06}.
Nickel has also been suggested to stretch the dose dependence of hardening or volume fraction~\cite{Eason13},
with Monte Carlo simulations finding the attractive Ni-Cu solute interaction both promotes nucleation~\cite{AlMotasem11,Wang17}
and slows down CIP growth by hindering the mobility of Cu clusters~\cite{Wang17}. The formation of stabilizing Ni shells on CIPs may also contribute to the retardation~\cite{Karkin19}.

\begin{figure*}
\begin{center}
\includegraphics[width=18.3cm]{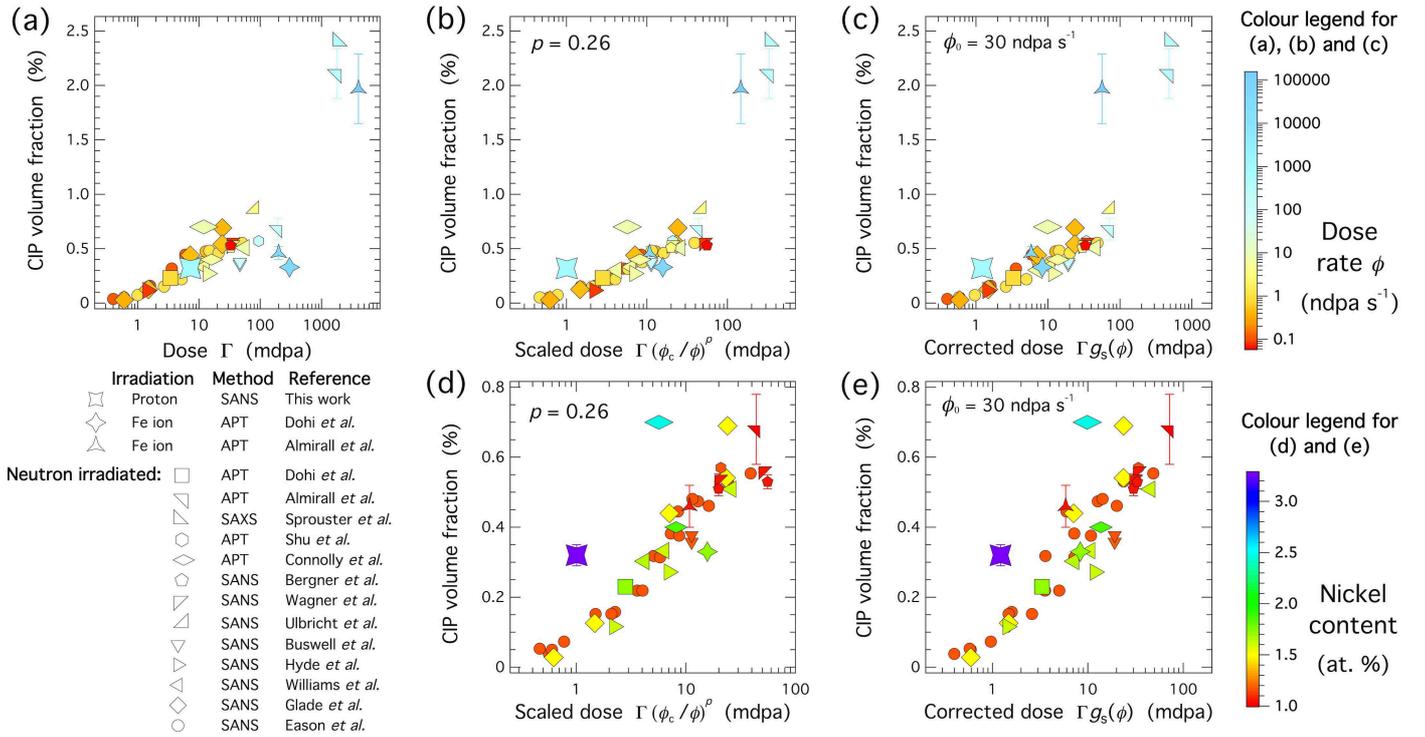}
\end{center}
\caption{\label{fig:scaling}(a) Comparison of CIP volume fraction in our 300$^\circ$C proton-irradiated specimen with results from the literature.
(b) and (c) Comparison of methods (see text Sec.~\ref{sec:volfracscaling}) to normalize the experimental dose $\Gamma$ for the effects
of dose rate $\phi$. (d) and (e) are zooms of (b) and (c) respectively and illustrate the effect of nickel.
References are Dohi {\it et al.}~\cite{Dohi10},
Almirall {\it et al.}~\cite{Almirall20}, Sprouster {\it et al.}~\cite{Sprouster16}, Shu {\it et al.}~\cite{Shu18a}, Connolly {\it et al.}~\cite{Connolly15}, Bergner {\it et al.}~\cite{Bergner08},
Wagner {\it et al.}~\cite{Wagner16}, Ulbricht {\it et al.}~\cite{Ulbricht05}, Buswell {\it et al.}~\cite{Buswell99}, Hyde {\it et al.}~\cite{Hyde01}, Williams {\it et al.}~\cite{Williams02},
Glade {\it et al.}~\cite{Glade06} and Eason {\it et al.}~\cite{Eason07}.}
\end{figure*}

We inspect our results more quantitatively by comparing measurements from the literature~\cite{Dohi10,Almirall20,Sprouster16,Shu18a,Connolly15,Bergner08,Wagner16,Ulbricht05,Buswell99,Hyde01,Williams02,Glade06,Eason07},
selecting those on RPV steels with Cu content between 0.19 and 0.33\,at.\% and Ni content $\geq 1$\,at.\%
in view of the leading effects of these elements (cf.\ Fig.~\ref{fig:schematic}). The highest Ni content in this comparison is that of our specimen (3.3\,at.\%).
In Fig.~\ref{fig:scaling}(a) we plot the volume fraction as a function of dose.
We have corrected for the different scattering contrasts used in some SANS studies for calculating $V$~\cite{Buswell99,Hyde01,Williams02}.
The data shown span dose rates between 0.06\,ndpa\,s$^{-1}$~\cite{Bergner08} and 150\,000\,ndpa\,s$^{-1}$~\cite{Almirall20};
irradiation temperatures between 225$^\circ$C~\cite{Connolly15} and 320$^\circ$C~\cite{Almirall20}; Mn content between 1.1~\cite{Shu18a,Almirall20} and 1.6\,at.\%~\cite{Glade06}.
An overlap with our 300$^\circ$C proton-irradiated result is seen.
However, considering the known effects of nickel
and that our steel has the highest Ni content on this plot (cf.\ colour scales of Figs.~\ref{fig:scaling}(d) and \ref{fig:scaling}(e)),
we expect our result should lead the trend instead of overlapping it.
Also the high dose-rate data (e.g.\ ion-irradiated points of Dohi {\it et al.}~\cite{Dohi10} or Almirall {\it et al.}~\cite{Almirall20})
lie to the right of the trend. This all indicates that dose rate effects are significant and should be corrected for.

\subsubsection{Dose rate effects on volume fraction}
\label{sec:volfracscaling}
We consider the effects of dose rate $\phi$ (or equivalently, of flux) at a given dose $\Gamma = \phi t$ (or fluence). Lower dose rates thus imply larger irradiation times to reach the same dose and
so there is an underlying effect of thermal ageing.
Monte Carlo simulations of neutron-irradiated bcc iron show this thermal ageing component becomes noticeable at dose rates below $\approx 0.1$\,ndpa\,s$^{-1}$~\cite{Soneda03}.
At higher dose rates the same study finds radiation damage is largely dose-rate independent up to $\approx 1000$\,ndpa\,s$^{-1}$~\cite{Soneda03}.
Above this threshold the recombination of vacancies produced in one cascade --- a process which takes a few picoseconds --- is not yet complete before another cascade occurs in the same region.
Higher doses are then required to achieve the same damage.
The threshold dose rates above depend on the mobility of vacancies and therefore upon their binding to solute atoms and microstructural defect sinks in real alloy systems.
Ballistic mixing --- the athermal mixing of solute atoms within cascades --- may also be anticipated to produce dose rate effects at high dose rates;
however these effects are limited by high sink densities~\cite{Soisson16} and are expected to be insignificant here, due e.g.\ to the solutes present in the steels considered~\cite{Shu18p}.
We might also expect ballistic mixing effects to be inconsequential for the small displacement cascades of our proton irradiation experiments.

Rate theory~\cite{Sizmann78,Mansur78} may be used to describe the recombination of vacancies and interstitial atoms and show how radiation becomes less effective in increasing the vacancy concentration
at higher dose rates. A simplified approach has been covered in several works~\cite{Odette05,Bergner08,Odette19}. We outline the main points here as they form a basis for various data analysis methods
and for further discussion.
The coupled pair of rate equations for the radiation-induced vacancy and interstitial fractions $x_\mathrm{v}$ and $x_\mathrm{i}$ are
\begin{equation} \dot{x}_\mathrm{v} = G_\mathrm{v} - R_\mathrm{f} x_\mathrm{i} x_\mathrm{v} - K_\mathrm{v} x_\mathrm{v}, \qquad
\dot{x}_\mathrm{i} = G_\mathrm{i} - R_\mathrm{f} x_\mathrm{i} x_\mathrm{v} - K_\mathrm{i} x_\mathrm{i} \label{eq:ratecouple} \end{equation}
where $\, \mathbf{\dot{ }} \,$ denotes the time derivative and $G_\mathrm{v} = G_\mathrm{i} = \xi \phi$ are vacancy and interstitial production rates in displacement cascades,
with $\xi$ the fraction of defects generated per dpa.
The reaction rate constants $K_\mathrm{v} = D_\mathrm{v} S$ and $K_\mathrm{i} = D_\mathrm{i} S$ describe consumption of defects at immobile sinks of fixed strength $S$.
$D_\mathrm{v}$ and $D_\mathrm{i}$ are vacancy and interstitial diffusivities.
In the steady state ($\dot{x}_\mathrm{v} = \dot{x}_\mathrm{i} = 0$), the recombination coefficient
$R_\mathrm{f} = 4 \pi r_\mathrm{f} (D_\mathrm{v} + D_\mathrm{i}) / \Omega_\mathrm{m}$~\cite{Waite57};
this is dominated by $D_\mathrm{i} \gg D_\mathrm{v}$ so $R_\mathrm{f} \approx 4 \pi r_\mathrm{f} D_\mathrm{i} / \Omega_\mathrm{m}$,
where $r_\mathrm{f}$ is the recombination radius.
The steady state solution satisfies the quadratic
$\alpha g_\mathrm{s}^2 + g_\mathrm{s} - 1 = 0$
where $g_\mathrm{s} = K_\mathrm{v} x_\mathrm{v} / G_\mathrm{v} = K_\mathrm{i} x_\mathrm{i} / G_\mathrm{i}$
is the fraction of vacancies and interstitials that escape recombination. Its positive root
\begin{equation} g_\mathrm{s} = \frac{(1 + 4 \alpha)^{1/2} - 1}{2 \alpha} \qquad \mbox{where} \quad \alpha = \frac{4 \pi r_\mathrm{f} \xi \phi}{\Omega_\mathrm{m} D_\mathrm{v} S^2} \label{eq:gs} \end{equation}

$\alpha$ is proportional to the dose rate $\phi$ with proportionality constant $1/\phi_0$ where $\phi_0 = \Omega_\mathrm{m} D_\mathrm{v} S^2 / 4 \pi r_\mathrm{f} \xi$.
Recombination dominates and $g_\mathrm{s} \approx \alpha^{-1/2}$ when $\alpha \gg 1$ in the high dose rate ($\phi \gg \phi_0$) regime.
In contrast $g_\mathrm{s}$ is unity in the sink dominated regime when $\alpha \ll 1$ at low dose rates $\phi \ll \phi_0$.
The growth of CIPs, which is driven by the radiation enhancement of diffusion, may therefore be expected to scale as $\phi^{-p}$ with $p=1/2$ at high dose rates
and to be independent of dose rate (i.e.\ $p=0$) at low dose rates.
This motivates one analysis method: power-law scaling of the experimental dose $\Gamma$, whereby
an effective scaled dose $\Gamma (\phi_\mathrm{c} / \phi)^p$ is defined to account for radiation-enhanced diffusion rate effects. $\phi_\mathrm{c}$ is the dose rate chosen to be invariant under scaling
and $p$ is usually found by inspection. For the data in Fig.~\ref{fig:scaling}(a) we find $p = 0.25$--$0.30$ gives satisfactory results
and show the result for $p=0.26$ in Fig.~\ref{fig:scaling}(b), where we take $\phi_\mathrm{c} = 0.45$\,ndpa\,s$^{-1}$ for consistency with Ref.~\cite{Eason07} and so that direct comparison with RPV surveillance data may be made.
It is remarkable that the volume fraction in Fig.~\ref{fig:scaling}(b) appears to scale satisfactorily over 6 orders of magnitude of dose rate,
from 0.06\,ndpa\,s$^{-1}$~\cite{Bergner08} to 150\,000\,ndpa\,s$^{-1}$~\cite{Almirall20}.

In Fig.~\ref{fig:scaling}(d) we focus on the modest dose region and show the nickel content of steels in this data set.
APT tip compositions are used where provided~\cite{Almirall20,Connolly15}.
The two alloys of greatest Ni content, i.e.\ our specimen (3.3\,at.\%) and that of Connolly {\it et al.}~\cite{Connolly15} (2.5\,at.\%)
clearly have CIP volume fractions that lie on a higher curve compared to the other alloys (that have 1.0--1.7\,at.\% Ni), as anticipated in view of the known nickel effects in synergy with copper
(see Fig.~\ref{fig:schematic} and last paragraph of Sec.~\ref{sec:introduction}).
Dose rate effects are also expected to depend somewhat on Ni content from previous experimental investigations~\cite{Was05,Odette05}.
Lower volume fractions at higher dose rates are clearly observed in SANS measurements on neutron irradiated RPV welds containing 0.2--0.6\,wt.\% Cu and 1.7\,wt.\% Ni, 
with $\phi$ varying between 0.08 and 7\,ndpa\,s$^{-1}$~\cite{Hyde01,Williams02}.
On the other hand, the irradiation hardening (and therefore CIP volume fraction) of Fe-Cu-Mn model alloys is observed to be independent of dose rate
over the large range probed (0.3--3000\,ndpa\,s$^{-1}$) with various irradiating particles (electrons, protons, neutrons)~\cite{Was05}.
Microstructure may also play a role. All the data in Fig.~\ref{fig:scaling} are from split-melt steels except the two points from Dohi {\it et al.}~\cite{Dohi10}; these are from a quenched ferritic steel.
Our $p$ values are in line with the range of values (0.15--0.40) reported for other Ni-bearing split-melt bainitic RPV steels~\cite{Was05,Odette19,Williams01},
whereas for simple quenched ferritic model alloys containing high Ni (1.6\,at.\%) $p = 0.5$ is reported~\cite{Was05}.
This may be due to a lower sink density and hence lower $S$ and $\phi_0$ in ferritic steels, placing them firmly in the recombination dominated regime.

A second method to correct for dose rate effects is to use Eq.~\eqref{eq:gs} directly~\cite{Wagner16,Odette19}. Presuming that a common $\phi_0$ describes
all data in Fig.~\ref{fig:scaling}, we find by inspection that
values of $\phi_0$ between 10 and 80\,ndpa\,s$^{-1}$ are most suitable and show the result for $\phi_0 = 30$\,ndpa\,s$^{-1}$ in Figs.~\ref{fig:scaling}(c) and \ref{fig:scaling}(e).
With $r_\mathrm{f} = 0.57$\,nm and $\xi = 0.4$, $\phi_0 \simeq 30$\,ndpa\,s$^{-1}$ implies $D_\mathrm{v} S^2 \simeq 7 \times 10^{12}$\,m$^{-2}$\,s$^{-1}$.
Neither $D_\mathrm{v}$ nor $S$ are known with great certainty.
Taking the sink strength $S$ as the dislocation density, values between $10^{13}$\,m$^{-2}$ and $10^{15}$\,m$^{-2}$ are expected in RPV steels (see e.g.\ \cite{Fujii13,Yoshida17,Kedharnath19,Shi19}).
With $S =  10^{14}$\,m$^{-2}$ for example, $D_\mathrm{v} \simeq 10^{-15}$\,m$^2$\,s$^{-1}$ at 300$^\circ$C,
consistent with suppressed $D_\mathrm{v}$
associated with higher effective vacancy migration energy
$\simeq 1.2$\,eV in ferrite compared to pure iron.
This last effect is due to the formation of vacancy-carbon complexes that need to dissociate before vacancies can diffuse~\cite{Fu08};
vC$_2$ complexes in particular are important at 300$^\circ$C~\cite{Jourdan11,Terentyev12,Konstantinovic17} (see Sec.~\ref{sec:carbon}).

\begin{figure}
\begin{center}
\includegraphics[width=8.1cm]{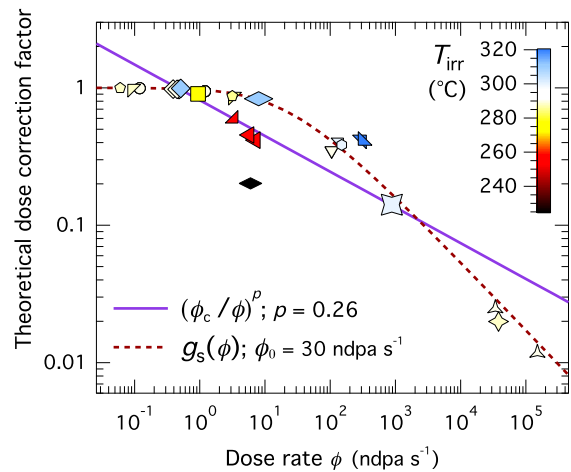}
\end{center}
\caption{\label{fig:escapefraction} Correction factors normalizing dose for dose rate effects, shown for the power-law scaling and $g_\mathrm{s}(\phi)$ methods used in Fig.~\ref{fig:scaling}
(solid and dashed lines respectively)
and for the solute trapping recombination (STR) model proposed by Odette {\it et al.}~\cite{Odette05} with parameters from Ref.~\cite{Odette19} (symbols). The sensitivity to irradiation temperature $T_\mathrm{irr}$
of the STR model is illustrated. Symbol key and data references are the same as in Fig.~\ref{fig:scaling}.}
\end{figure}
It is worth exploring dose rate effects and deriving normalization methods since this may reveal microscopic mechanisms underpinning CIP growth.
Bearing in mind that the $\phi^{-1/2}$ dependence emerging from the simple vacancy-interstitial recombination model is a result of its quadratic kernel,
we note that if a $\phi^{-1/n}$ dependence is evidenced experimentally, then at least $n$ distinct species must be significant active agents in the diffusion of point defects and in CIP growth.
Indeed the addition of vacancy-solute traps to the vacancy-interstitial recombination model is suggested to yield a $\phi^{-1/3}$ dependence at low dose rates~\cite{Odette05}.
In Fig.~\ref{fig:escapefraction} we compare dose correction factors for the power-law scaling (solid line in Fig.~\ref{fig:escapefraction}) and $g_\mathrm{s}(\phi)$ (dashed line) methods.
Both result in the two steels with highest nickel concentrations lying above the other alloys in CIP volume fraction (Fig.~\ref{fig:scaling}),
but on balance the power-law scaling appears more satisfactory in correcting the CIP volume fraction for dose rate effects.
The difference between correction methods should be most noticeable at the extremes in dose rate (cf.\ Fig.~\ref{fig:escapefraction}). There we see the power-law scaling prevails:
it both better normalizes the low dose-rate data of Eason {\it et al.}~\cite{Eason07} (orange circles in Figs.~\ref{fig:scaling}(d) and Figs.~\ref{fig:scaling}(e)),
while upholding a more sensible normalization of the high dose-rate ion-irradiation data of Almirall {\it et al.}~\cite{Almirall20}
(blue symbols in the right halves of Figs.~\ref{fig:scaling}(b) and Figs.~\ref{fig:scaling}(c)).
By contrast the $g_\mathrm{s}(\phi)$ adjustment returns a wider spread in the data at modest doses (cf.\ Figs.~\ref{fig:scaling}(d) and \ref{fig:scaling}(e))
and misorders the pair of ion-irradiated and neutron-irradiated points from Almirall {\it et al.}~\cite{Almirall20} at $\approx 70$\,mdpa in $g_\mathrm{s}(\phi)$-adjusted dose (Fig.~\ref{fig:scaling}(c))
that have very different volume fractions but are in fact irradiations of the same steel,
so these volume fractions should be monotonically increasing with effective dose.
For completeness in Fig.~\ref{fig:escapefraction} we also compare the solute trapping recombination (STR) model of Odette {\it et al.}~\cite{Odette05},
taking the concentration of available traps as the total content of vacancy-binding solutes Cu, Ni, Mn and Si~\cite{Ohnuma09,Messina14} (symbols in Fig.~\ref{fig:escapefraction}).
We use the approximation and parameter values detailed in Ref.~\cite{Odette19} but increase the sink strength to $10^{15}$\,m$^{-2}$; this facilitates comparison with the other two models
by shifting the crossover between sink- and recombination-dominated regimes to higher dose rates.
From Fig.~\ref{fig:escapefraction} we see the STR model gives similar corrections to the $g_\mathrm{s}(\phi)$ method with common $\phi_0$,
but is less satisfactory as it appears to over-correct for the effects of irradiation temperature.
An improved rate theory model might account for vacancy-carbon complexes and the non-Arrhenius vacancy diffusivity that results~\cite{Fu08}.
Moreover it should describe trapping of self interstitial atoms e.g.\ in Fe-Mn dumbbells~\cite{Messina20}. Around half of interstitials generated in cascades are clustered~\cite{Marinica12},
so it would not be surprising if triplets and larger clusters~\cite{Marinica12,Becquart18,Whiting19} were also significant in dose rate effects.
It will be intriguing to discover in future work which four distinct species are instrumental in manifesting the $\phi^{-1/4}$ scaling ostensively supported by experiments on Ni-bearing RPV steels.

\begin{figure*}
\begin{center}
\includegraphics[width=18.3cm]{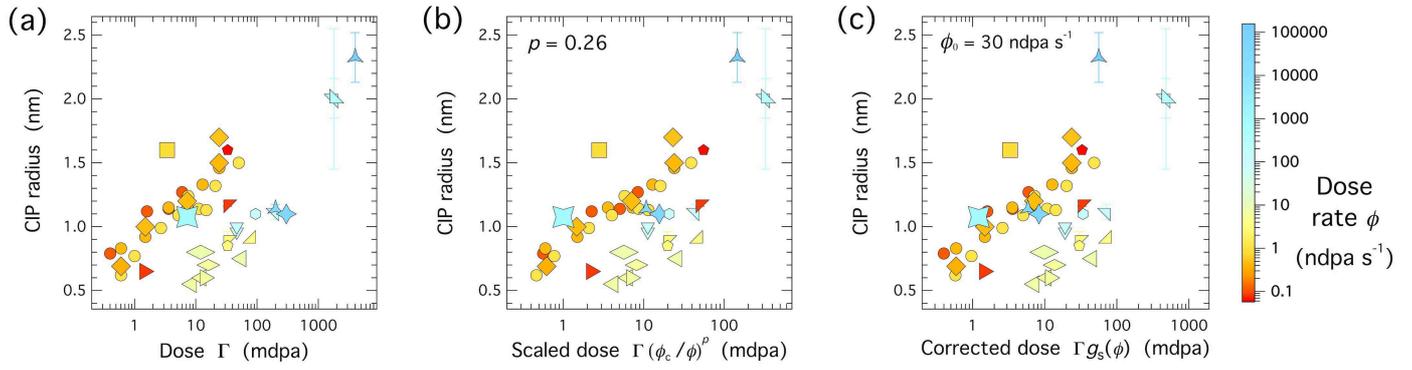}
\end{center}
\caption{\label{fig:radscaling}Comparison of CIP radius in our 300$^\circ$C proton-irradiated specimen with results from the literature.
Effects of employing the two methods from Fig.~\ref{fig:scaling} for normalizing the experimental dose $\Gamma$ are shown in (b) and (c).
See Fig.~\ref{fig:scaling} legend for symbol key and data references.}
\end{figure*}

\subsubsection{Dose rate effects and possible influence of Ni, Mn and P on radius}
\label{sec:radius}
In Fig.~\ref{fig:radscaling}(a) we compare the CIP radius in our 300$^\circ$C proton-irradiated specimen with the same data set used in Fig.~\ref{fig:scaling}
consisting of similar Cu and high Ni steels from the literature.
We see higher dose rates tend to produce smaller CIPs for a given dose, as expected from the literature (e.g.\ \cite{Wagner16,Shu18p}).
Since this effect on radius is visibly more pronounced than that on volume fraction (cf.\ Fig.~\ref{fig:scaling}(a)), this means the number density of CIPs increases with dose rate~\cite{Wagner16}.
This outcome of more numerous and smaller CIPs is reminiscent of the known effects of nickel and manganese~\cite{Miller07,Miller03,Glade05,Meslin10,Shu18m,Shu18p}
(see Sec.~\ref{sec:nickel})
and leads us to conjecture that higher dose rates similarly suppress the CIP-matrix interface energy or the work needed to form a critical nucleus.

In Figs.~\ref{fig:radscaling}(b) and \ref{fig:radscaling}(c) we see the methods to normalize for dose rate effects that worked well for CIP volume fraction
appear, at first glance, to fail for CIP radius.
The surviving variations in the data after normalization are difficult to justify, even with sophisticated cluster dynamics modelling
where dose-rate scaling is extraneously enforced~\cite{Mamivand19}.
We suggest another factor plays a key role in determining CIP sizes and in the effects of dose rate therein.
\begin{figure}
\begin{center}
\includegraphics[width=8.1cm]{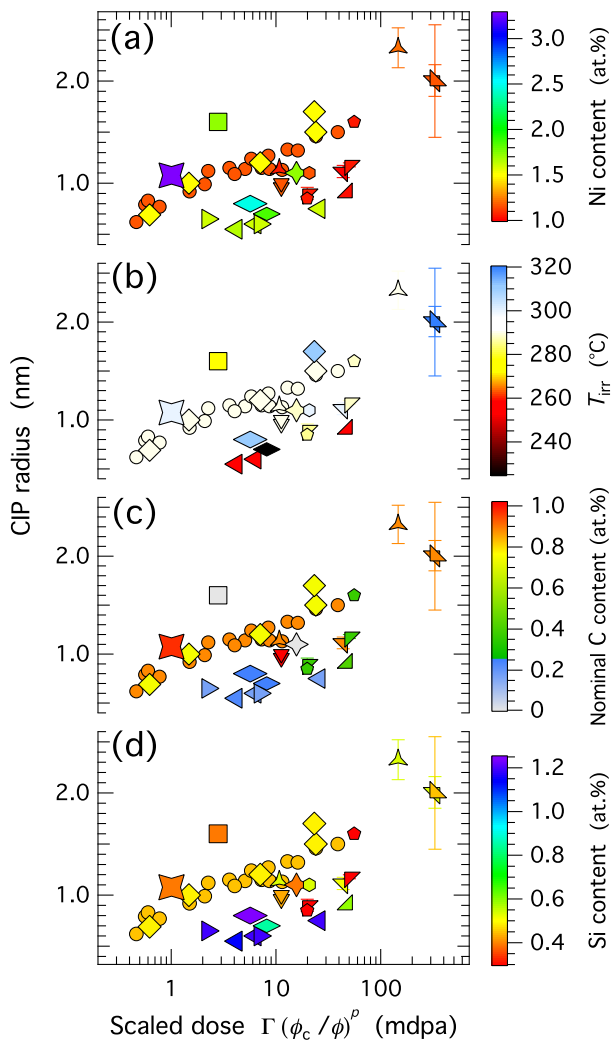}
\end{center}
\caption{\label{fig:radfactors} Reproductions of CIP radii data from Fig.~\ref{fig:radscaling}(b) illustrating relative effects of (a) nickel content,
(b) irradiation temperature, (c) carbon content and (d) silicon content. See Fig.~\ref{fig:scaling} legend for symbol key and data references.}
\end{figure}
To reveal this hidden factor, in Fig.~\ref{fig:radfactors} we repeat the plot of Fig.~\ref{fig:radscaling}(b), highlighting other factors in place of dose rate.
In Fig.~\ref{fig:radfactors}(a) we plot the Ni content and see no systematic variation.
We also do not find any systematic variations of either CIP volume fractions or sizes with manganese or phosphorus (not shown).
The P content varies between 0.004\,at.\%~\cite{Dohi10} and 0.1\,at.\%~\cite{Buswell99} for this data set.
The lack of any discernible dependence of CIP sizes on Ni or Mn is surprising given that both are expected to result in smaller CIPs~\cite{Miller07,Miller03,Glade05,Meslin10,Shu18m,Shu18p}.
The Mn content varies only between 1.1 and 1.6\,at.\% for this data set, however. The lack of systematic variation of CIP radii with Ni content (Fig.~\ref{fig:radfactors}(a))
suggests that the nickel effect on CIP sizes is saturated at high Ni contents $\gtrsim 1$\,at.\%.

\subsection{Irradiation temperature and shift for surrogate ion experiments}
\label{sec:temperature}
Higher irradiation temperatures $T_\mathrm{irr}$ are known to lead to reduced hardening~\cite{Williams19,Debarberis05} with a corresponding decrease in CIP volume fraction.
This is consistent with our observations of a $\approx 20$\% lower CIP volume fraction following proton irradiation at 400$^\circ$C compared to that at 300$^\circ$C (cf.\ Table~\ref{tab:SANS}).
However, the effect on CIP volume fraction is only slight for the literature data set shown in Fig.~\ref{fig:scaling} for which $T_\mathrm{irr}$ spans from
225$^\circ$C~\cite{Connolly15} to 320$^\circ$C~\cite{Almirall20}.
Higher $T_\mathrm{irr}$ is also thought to increase CIP sizes (see e.g.\ \cite{Glade06})
and indeed we observe larger CIPs at proton irradiation temperatures of 400$^\circ$C compared to 300$^\circ$C (cf.\ Table~\ref{tab:SANS}).
We also see in Fig.~\ref{fig:radfactors}(b) that the trend of larger CIP radii at higher $T_\mathrm{irr}$ is mostly borne out in the literature data set
but there are notable exceptions: for example the 310$^\circ$C point from Connolly {\it et al.}~\cite{Connolly15} (blue rhombus at $\simeq 6$\,mdpa scaled dose in Fig.~\ref{fig:radfactors}(b))
and the 276$^\circ$C point from Dohi {\it et al.}~\cite{Dohi10} (yellow square in Fig.~\ref{fig:radfactors}(b)).
In Sec.~\ref{sec:carbon} we see that these exceptions can be explained by a strong dependence of CIP sizes on carbon content.

Comparing our 300$^\circ$C and 400$^\circ$C proton irradiations to neutron irradiations at $\approx 300^\circ$C from the literature,
we find overall that 300$^\circ$C proton irradiation, with a correction of dose for the accelerated dose rate, yields CIPs similar to those engendered by neutrons,
with excellent agreement with literature data in CIP volume fractions (Fig.~\ref{fig:scaling}), sizes (Figs.~\ref{fig:radscaling} and \ref{fig:radfactors})
and composition (Fig.~\ref{fig:ternary}).
On the other hand, a shift of proton irradiation temperature to 400$^\circ$C causes CIP sizes and composition to diverge away from $\approx 300^\circ$C neutron literature data,
as is readily illustrated, for example, for CIP compositions by the ternary diagram (Fig.~\ref{fig:ternary}) where
the measured $M/N$ ratio contours for our 400$^\circ$C proton-irradiated specimen fail to overlap with APT results from the literature.

A zero shift in irradiation temperature is evidently desirable when simulating precipitation in reactor steels using surrogate particles,
if precipitate volume fractions, sizes and compositions are to be reproduced.
There are several reasons for this.
To begin with, in Cu-containing steels, the proximity of Cu content and reactor steel operating temperatures to the Cu solubility limit
means that the degree of Cu supersaturation is a strong function of temperature; for example,
a 1\% change in absolute temperature at 573\,K produces a $\approx 20$\% change in Cu supersaturation~\cite{Salje77}.
The stability of vacancy-carbon complexes is also a strong function of temperature~\cite{Terentyev12}.
A zero shift in irradiation temperature is therefore desirable
since we suggest shortly in Sec.~\ref{sec:carbon} that dissolved carbon plays a key role in determining precipitate sizes in conjunction with dose rate.
Further, irradiation temperature and dose rate are known from experiments to have qualitatively dissimilar effects on precipitate development:
higher irradiation temperatures and higher dose rates both result in suppressed CIP volume fractions,
but higher irradiation temperatures tend to increase CIP sizes,
whereas higher dose rates tend to produce smaller CIPs for a given dose.
Temperature cannot therefore be used to compensate for dose rate effects as done e.g.\ in austenitic stainless steels to reproduce interfacial segregation.

\subsection{Potential effects of carbon}
\label{sec:carbon}
In Fig.~\ref{fig:radfactors}(c) we show the nominal carbon content, or analysed average carbon content where available,
of steels in the data set used in Fig.~\ref{fig:scaling}.
Ignoring the two grey points with $\simeq 0.005$\,at.\% C of Dohi~{\it et al.}~\cite{Dohi10},
we see an overall trend towards smaller CIP radii at lower nominal carbon content for a given dose.
The variation due to carbon appears at least as significant as that due to dose rate (cf.\ Fig.~\ref{fig:radscaling}),
so normalizations accounting only for dose rate are unlikely to succeed.
The strong dependence of CIP sizes on C content may surpass the effect of irradiation temperature, explaining the exceptions to expected $T_\mathrm{irr}$ effects noted above in Sec.~\ref{sec:temperature}.

The potential roles played by carbon are manifold. It determines the microstructure depending on heat treatment.
The two grey points of Dohi~{\it et al.}~\cite{Dohi10} are simply quenched and have ferritic microstructures~\cite{Dohi10},
but all other steels in Fig.~\ref{fig:radfactors}(c) have carbon contents well above the ferrite solubility limit and are expected to have appreciable carbide fractions.
Carbide-ferrite interfaces are ostensibly effective defect sinks, particularly for dumbbells and vacancy-solute pairs composed of carbon-binding solutes.
We noted in Sec.~\ref{sec:volfracscaling} that the low sink density in simple ferritic alloys
leads to recombination-dominated dose-rate scaling of CIP volume fractions~\cite{Was05}.
Looking for possible microstructural effects on CIP sizes, in Fig.~\ref{fig:radfactors}(c) we find the blue and green points with lower nominal carbon contents $\lesssim 0.4$\,at.\% are all welds,
whereas the red, orange and yellow points with higher nominal carbon contents $\geq 0.7$\,at.\% are split-melt plates or forgings.
At first glance this distinction seems to divide the data set into two groups, with generally smaller CIPs in welds.
However the weld data from Bergner {\it et al.}~\cite{Bergner08} (top green pentagon in Fig.~\ref{fig:radfactors}(c))
demonstrate that welds can manifest larger CIPs.
Additionally, intermediate-sized CIPs falling in the apparent gap between welds and split-melt steels are observed in welds containing more carbon (up to 0.6\,at.\%)
than those in our data set~\cite{Miller07,Carter01}; they are not plotted in Figs.~\ref{fig:scaling}--\ref{fig:radfactors} as
their Cu and Ni contents fall outside the bounds delimiting our data set (see Sec.~\ref{sec:nickel}).

Welds tend to have lower carbon content and also higher silicon content than the base metal, depending on the weld flux.
The 7 blue weld points of lowest C content $\leq 0.25$\,at.\% in Fig.~\ref{fig:radfactors}(c) indeed have the highest Si contents $\geq 0.85$\,at.\% of our data set,
as illustrated in Fig.~\ref{fig:radfactors}(d).
However 4 of the 5 green weld points of C content $\simeq 0.4$\,at.\% in Fig.~\ref{fig:radfactors}(c) have the lowest Si content ($0.3$\,at.\%) of our data set.
As both blue and green points in Fig.~\ref{fig:radfactors}(c) generally have smaller CIPs for a given dose, we deduce that any silicon effects are secondary to carbon or microstructural effects.
This is corroborated by literature reporting identical CIP radii following similar neutron irradiations of RPV welds of disparate Si content (0.4\,at.\% and 1.2\,at.\%)
but comparable C content (0.6\,at.\% and 0.4\,at.\%)~\cite{Carter01}.

Further studies, in particular on lower carbon model steels, are needed to separate possible direct effects of carbon
from microstructural effects particular and common to welds that may promote CIP nucleation.
Further characterizations of matrix carbon levels and how carbon is distributed in RPV steels are required.
For the time being we construct a picture of how carbon effects might work, positing that higher nominal contents above the ferrite solubility limit lead somewhat monotonically
to greater carbide fractions and also to higher residual carbon levels in the ferrite matrix.
This premise is not inconceivable given all welds and split-melt steels in our data set have comparable post-weld stress-relief or tempering treatments of
several hours at temperatures 600$^\circ$C--650$^\circ$C.
Greater carbide fractions will mop up carbide-forming solutes like manganese and diminish their levels in the matrix.
Recalling that the known leading effect of manganese on CIPs is to restrict their sizes,
we expect this explains some of the trend evinced in Fig.~\ref{fig:radfactors}(c),
with less Mn remaining in the matrix reducing its size-limiting effect on CIPs,
resulting in larger CIPs at higher nominal carbon contents.
There are, however, data such as the pair of green pentagons in Fig.~\ref{fig:radfactors}(c) from Bergner {\it et al.}~\cite{Bergner08}
that are measured on the same steel irradiated to the same dose but at different dose rates, with very different CIP radii produced.
There must therefore also be a significant compound or synergistic effect of carbon with dose rate.

Due to the strong attraction between vacancies and interstitial carbon, vacancy-carbon complexes are expected to form under irradiation~\cite{Ohnuma09,Jourdan11,Terentyev14}.
We anticipate larger complexes like vC$_2$~\cite{Jourdan11,Terentyev12,Konstantinovic17} will be significant in dose rate effects,
since at irradiation temperatures $\approx 300^\circ$C these lie on the threshold of stability as evidenced by positron annihilation~\cite{Terentyev14}
and resistivity recovery~\cite{Takaki83} experiments.
Their long metastable lifetimes will span the interval between displacement cascades occurring in the same sample region,
with the overlap extent depending on dose rate.
Furthermore these lifetimes will strongly depend on irradiation temperature around 300$^\circ$C and may well drive the known effects of irradiation temperature on CIP volumes and sizes.
With effective migration energies of $\simeq 1.6$\,eV~\cite{Barouh15}, vC$_2$ complexes are immobile and should act as effective sinks for interstitials,
such as Fe-Mn dumbbells that dominate the transport of Mn~\cite{Meslin13,Messina20}.
There will also be sink strengthening in the vacancy-solute transport channel due to dissolved carbon binding with vacancy-solute pairs to form immobile triplets~\cite{Bakaev14,LiuXing14}.
To explain the compound effect of carbon and dose rate on CIP radii, a net bias in sink strengthening caused by carbon
toward the interstitial channels is required i.e.\ trapping by carbon and vacancy-carbon complexes of Fe-Mn dumbbells should outweigh that of vacancy-solute pairs.
This would subsequently result in a relative easing of the suppression, due to recombination at higher dose rates, in interstitial concentrations,
so at higher dose rates Fe-Mn dumbbell concentrations will be relatively greater compared to free vacancy and vacancy-solute concentrations that drive initial CIP growth by Cu transport.
This effective increase in manganese effect would lead to smaller CIPs at higher dose rates, as seen in Fig.~\ref{fig:radscaling}.

In Sec.~\ref{sec:volfracscaling} we argued that a minimum of 4 distinct defect species are required in rate theory to reproduce the power-law scaling of
dose for dose rate effects evinced by the experimental data.
Fe-Mn dumbbells, carbon-containing vacancy complexes and carbon-containing solute clusters are clear candidates to include in future rate theory work.
This would then couple with cluster dynamics modelling, for example by quantifying the effects of dose rate on interfacial energy,
while atomistic simulations could explore possible changes in solute kinetics in the vicinity of precipitates.

\section{Conclusions}
In this study high Ni, Cu-containing model steel specimens were irradiated with 5\,MeV protons to average doses of 7.2\,mdpa at dose rates 10$^4$ times greater than those experienced
due to fission neutrons by reactor pressure vessels (RPVs) in light-water reactor operation.
Small-angle neutron scattering (SANS) was used to characterize the resulting copper-initiated precipitates (CIPs), with the following principal results:
\begin{itemize}
\item Following proton irradiation at 300$^\circ$C, CIPs follow a Gaussian size distribution of mean radius 1.1\,nm and standard deviation 0.2\,nm.
The magnetic-to-nuclear SANS scattering ratio $M/N = 2.4 \pm 1.5$.
\item Following proton irradiation at 400$^\circ$C, CIPs are significantly larger than at 300$^\circ$C and give a higher $M/N$ ratio on average. The CIP size distribution is positively skewed:
it can be well described either by a single log-normal distribution or by a distribution with two size components. The latter is consistent with a picture
of small Cu cores surrounded by larger appendages rich in Ni, Mn and Si.
\end{itemize}
To establish further conclusions our SANS results were compared with data from the literature on high Ni, Cu-bearing RPV steels. We find:
\begin{itemize}
\item For CIP compositions in these steels, the measured SANS $M/N$ ratio is most sensitive to the Mn content of CIPs, is sensitive to Cu and Ni but not to these elements separately,
and is least sensitive to Si.
For our 300$^\circ$C proton-irradiated specimen, a CIP composition of $\approx 11$\,at.\% Cu, $\approx 47$\,at.\% Ni, $\approx 22$\,at.\% Mn and $\approx 20$\,at.\% Si is consistent with the measured $M/N$
and is argued to be the likely CIP composition after considering the effects of irradiation versus thermal ageing on CIP compositions and that identical steels
were thermally aged and characterized using atom probe tomography (APT) by Zelenty {\it et al.}~\cite{Zelenty16}.
\item A power-law scaling of the experimental dose for dose rate effects, with exponent 0.25--0.30, is found to acceptably align CIP volume fractions over 6 orders of magnitude in dose rate.
\item Scaling for dose rate effects does not work satisfactorily for CIP sizes and we suggest other effects are also dominant here.
We identify carbon content as playing a significant role in determining CIP sizes, in conjunction with dose rate.
\item 300$^\circ$C proton irradiation satisfactorily reproduces the CIP volume fraction, size and composition expected after irradiation by fission neutrons
at 300$^\circ$C reactor operating temperatures. 400$^\circ$C proton irradiation yields larger CIP sizes and CIP compositions with more Cu and/or Ni on average.
We therefore conclude a zero shift in irradiation temperature is desirable for simulating precipitation in reactor steels using surrogate particles at accelerated dose rates.
\end{itemize}
To close, we have demonstrated the partnership of proton irradiation and SANS is invaluable in investigations of radiation effects,
as proven in this study of precipitation in reactor steels.
MeV protons penetrate to a sufficient depth in metals so as to make bulk volume techniques like SANS feasible.
By characterizing the bulk of specimens, SANS yields precise assessments of volume fractions and size distributions of hardening precipitates,
complementing APT studies that provide compositional details but are limited to small probe volumes.
The partnership of proton irradiation and SANS facilitates precise studies of radiation effects at accelerated dose rates, enabling new research avenues to reveal underlying radiation damage mechanisms,
such as systematic studies of dose rate effects. It also clears a much-needed~\cite{Knaster16,GonzalezDeVicente17} irradiation--characterization route for future studies of fusion neutron effects.

\section*{Data availability}
The raw data required to reproduce these findings are available upon request.

\section*{Declaration of competing interest}
The authors declare that they have no known competing financial interests or personal relationships that could have appeared to influence the work reported in this paper.

\section*{CRediT authorship contribution statement}
\textbf{Mark Laver:} Conceptualization, Validation, Formal analysis, Investigation, Visualization, Writing - original draft, Writing - review \& editing, Supervision.
\textbf{Brian J.\ Connolly:} Conceptualization, Validation, Investigation, Writing - review \& editing, Supervision.
\textbf{Christopher Cooper:} Investigation, Writing - review \& editing.
\textbf{Joachim Kohlbrecher:} Investigation, Resources, Data curation, Writing - review \& editing.
\textbf{Stavros Samothrakitis:} Validation, Formal analysis, Writing - review \& editing.
\textbf{Keith Wilford:} Conceptualization, Resources, Writing - review \& editing.

\section*{Acknowledgements}
This work is based on experiments performed at the Swiss spallation neutron source SINQ, Paul Scherrer Institute, Villigen, Switzerland.
The authors wish to thank Rolls-Royce for providing the material used in this study,
the MC40 cyclotron team at the University of Birmingham for assistance with proton irradiations,
and Jonathan Young, Robert Arnold, Lionel Porcar, Lisa DeBeer-Schmitt and Uwe Keiderling for their work and support in complementary
microstructural characterizations of other steel specimens.
We acknowledge helpful comments from Nick Riddle and
the contributions of successive undergraduate nuclear engineering project students including Joe S.\ Roberts, Ela Young, Tom Heneghan,
Sanchit J.\ Nardekar and Patrick Roberts.
We sincerely thank the two anonymous reviewers whose comments helped improve the clarity of this manuscript.
This research did not receive any specific grant from funding agencies in the public, commercial, or not-for-profit sectors.

\bibliography{2021LaverBibliography}

\end{document}